# Perceptual Compensation of Ambisonics Recordings for Reproduction in Room

Ali Fallah[1], Shun Nakamura[1], Steven van de Par[1]

*Department of Medical Physics and Acoustics, Carl von Ossietzky University of Oldenburg, Cluster of Excellence – Hearing4all, Oldenburg, Germany*

## Abstract

Ambisonics is a method for capturing and rendering a sound field accurately, assuming that the acoustics of the playback room does not significantly influence the sound field. However, in practice, the acoustics of the playback room may lead to a noticeable degradation in sound quality. We propose a recording and rendering method based on Ambisonics that utilizes a perceptually-motivated approach to compensate for the reverberation of the playback room. The recorded direct and reverberant sound field components in the spherical harmonics (SHs) domain are spectrally and spatially compensated to preserve the relevant auditory cues including the direction of arrival of the direct sound, the spectral energy of the direct and reverberant sound components, and the Interaural Coherence (IC) across each auditory band. In contrast to the conventional Ambisonics, a flexible number of Ambisonics channels can be used for audio rendering. Listening test results show that the proposed method provides a perceptually accurate rendering of the originally recorded sound field, outperforming both conventional Ambisonics without compensation and even ideal Ambisonics rendering in a simulated anechoic room. Additionally, subjective evaluations of listeners seated at the center of the loudspeaker array demonstrate that the method remains robust to head rotation and minor displacements.

## I.    INTRODUCTION

An objective of an audio rendering system is to capture an acoustic scene, such as a concert hall performance, and reproduce it in another room in a way that listeners in the playback room perceive the sound field as closely matching the original performance. Theoretically, methods such as wave field synthesis (WFS) (Berkhout, 1988) and Ambisonics (Gerzon, 1973; Daniel, 2000), allow for the perfect reproduction of a recorded sound field. However, both methods have a limitation where the reproduction room ideally needs to have boundaries that are free of reflections. In addition, spacing of loudspeakers must be very dense in order to avoid spatial aliasing effects. In practice, distortions arising from the use of a limited number of loudspeakers and reverberation in the playback room can significantly reduce the accuracy of sound field reproduction. Additionally, loudspeakers typically exhibit non-flat frequency responses. Reproduction of a recorded reverberant sound field on loudspeakers that are themselves placed in a reverberant playback environment is known as Room-in-Room (RinR) sound reproduction, which has been discussed to alter the perceived sound field by introducing additional reflections, reverberation, and potential coloration, thereby distorting the accuracy of the original recording and reducing spatial fidelity (Merimaa and Pulkki, 2005). Additionally, most currently available recording microphones support only up to 4th or 6th order Ambisonics (mh-acoustics, 2023; 2025). At these orders, accurate sound-field reconstruction in the high-frequency range is confined to a small area, referred to as the sweet spot (Ahrens and Spors, 2009).

In the domain of Wave Field Synthesis, various approaches have been developed to compensate for the effects of the reproduction room. A method based on WFS theory and plane-wave decomposition was proposed to compensate for the room reflection over a large area by synthesizing the desired sound field while explicitly modeling and canceling the effects of early reflections (Spors *et al.*, 2003). Adaptive Wave Field Synthesis (AWFS) methods (Gauthier and Berry, 2006; Stefanakis *et* *al.*, 2010) and similar approaches treat dereverberation as an inverse problem, aiming to equalize the playback system's transfer function at multiple points in order to minimize reproduction errors at these control positions. However, such multi-point sampling and equalization methods can lead to poor equalization at positions outside the control points (Fielder, 2001).

In the Ambisonics domain, hardware solutions exist to control the reverberant field, such as using directive loudspeakers to reduce the energy of reflections (Poletti *et al.*, 2010). A method for two-dimensional room equalization over an extended area, employing cylindrical harmonics to model room reflections, has been proposed (Betlehem and Abhayapala, 2005). However, a drawback of this system is the need for a very large number of equalization filters. A modified version of this method for three-dimensional (3D) space was introduced (Lecomte *et al.*, 2018). In contrast to the approach in (Betlehem and Abhayapala, 2005) , this method introduces a new filtering matrix that reduces the number of equalization filters and simplifies the matrix inversion process. In multizone reproduction systems, intensity-matching techniques have been proposed to align the energy-flux direction of the reproduced field with the desired direction, thereby enhancing spatial cues across larger listening areas (Zuo *et al.*, 2021).

Inverse filtering in Ambisonics reproduction faces several challenges, particularly in room environments. It suffers from robustness issues and is highly sensitive to room acoustics, as variations in listener or loudspeaker positions and time-varying room reflections can undermine its effectiveness (Lecomte *et al.*, 2018). The inverse filtering also carries the risk of amplifying noise and errors particularly at higher frequencies due to measurement inaccuracies (Zhang *et al.*, 2008). Computational complexity also adds to the difficulty, as creating and implementing real-time filters for high-order systems requires significant resources (Hold *et al.*, 2023). Moreover, listener movement outside the calibrated area or in dynamic environments can make the filtering ineffective (Mouchtaris *et al.*, 2000). Furthermore, inverse filtering may introduce perceptual artifacts, such as over-correction, resulting in the reproduction of an unnatural sound field that undermines the immersive experience(Norcross *et al.*, 2004). Finally, these methods cannot compensate for the inherent problems in low-order Ambisonics recording and reproduction (Moreau *et al.*, 2006).

Complementary to physically based methods like Wave Field Synthesis (WFS) and Ambisonics, perceptually motivated approaches focus on delivering a subjectively accurate sound experience, even if the exact sound field is not precisely recreated. One such approach is Directional Audio Coding (DirAC) (Pulkki, 2007), which decomposes sound into directional and spatially diffuse components based on a diffuseness criterion in the time-frequency domain. These components are then processed and reproduced through separate signal paths. The direct sound, retaining its directional cues, is estimated in short time frames per frequency band, and rendered using the Vector Base Amplitude Panning (VBAP) method  (Pulkki, 1997). Parallel to that, the diffuse sound is rendered through multiple loudspeakers using decorrelation techniques to create a realistic, spatially diffuse sound field. Moreover, a modified version of DirAC has been proposed, utilizing spherical array recordings to be able to better extract multiple simultaneous sound sources (Politis *et al.*, 2015). The DirAC approach does not account for the reverberation of the reproduction room.

Instead of suppressing the reverberation of the playback room, which leads to robustness issues and practical challenges with inverse filtering, a method for recording and rendering was proposed in (Grosse and van de Par, 2015) that perceptually compensates for the effects of the playback room. In this approach, a near-source microphone is used to capture the direct sound, while two distant microphones are employed to record the reverberant sound. The processed direct sound is then reproduced using two loudspeakers positioned in front of the listener. The processed reverberant sound is reproduced using two distant dipole loudspeakers, oriented so that their directional patterns predominantly point away from the listener, with the aim of acoustically only exciting the reverberant sound field without creating noticeable direct sound-path components. A perceptually-motivated optimization procedure is employed, where the energies of the direct and reverberant sounds are separately compensated in each frequency band. This is achieved by equalizing the reproduced RinR impulse responses to ensure they match for the listener in both the recording and playback rooms. The optimization process also involves compensating for Interaural Coherence (IC) (Faller and Merimaa, 2004) between the left and right ear signals to preserve the perceived spatial diffuseness of the sound field in the reproduction room. To design the compensation filters, a dummy head is placed in the recording room as a reference, while another dummy head is situated in the playback room for optimization purposes. By incorporating the playback room in the optimization process, the acoustics of the playback room is perceptually compensated.

Relatively few studies in the literature address compensating for the reverberation of playback rooms in the Ambisonics domain. This study proposes a perceptually motivated compensation method in the Ambisonics domain, inspired by the primary approach in (Grosse and van de Par, 2015). However, instead of using distributed microphones, a conventional Ambisonics recording is employed using a single spherical microphone placed at the listener position. In the proposed method, the direct and reverberant components are separated after recording, then perceptually compensated and reproduced through separate signal paths.

The paper is organized as follows: Section II explains the signal model and audio recording approach, Section III covers the signal filtering and reproduction process, Section IV describes the optimization procedure, and Section V presents an evaluation of the algorithm. Finally, Section VI concludes the paper with a summary and suggestions for future work.

## II. PROPOSED METHOS

The main idea is illustrated in FIG.1. As can be seen, the spatial sound field is captured in the recording venue using a rigid spherical microphone array.

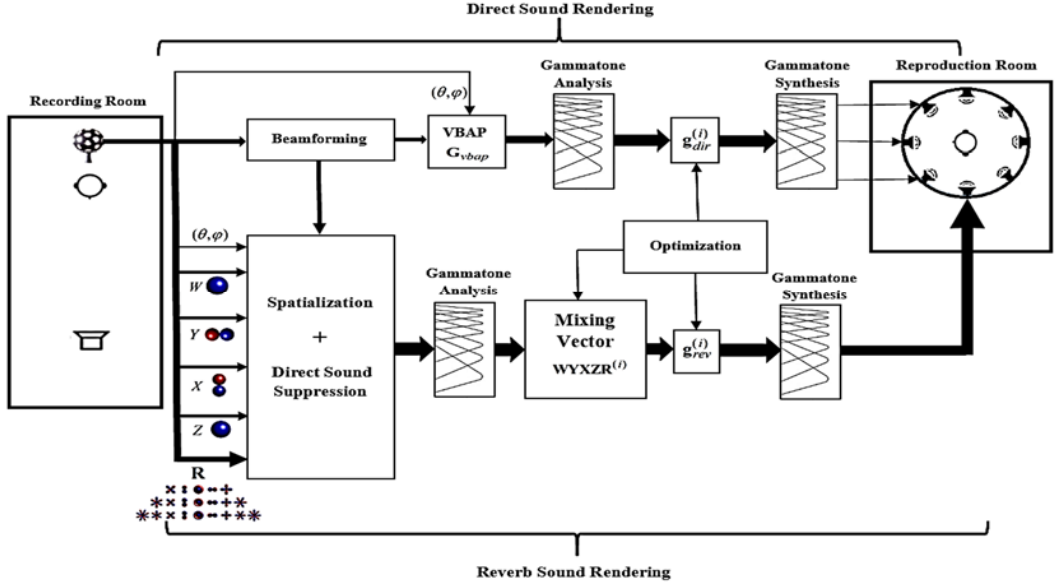

FIG.1. A block diagram of the Ambisonics-based recording, compensation and rendering approach. The sound is recorded using a microphone array and then separated into the direct signal path (upper path) and reverb signal path (lower path). Both the direct and reverb signals are compensated using Gammatone filterbank analysis and synthesis, before reproduction in the playback room. The direct sound is captured via beamforming and after being mapped onto three VBAP loudspeakers with the corresponding gains $\mathbf{G_{vbap}}$, is filtered using the gains of the Gammatone filters $g_{dir}^{(i)}$ to optimize the energy of the direct sound before playback through the VBAP loudspeakers. All Ambisonic channels up to $4^{th}$ order can be used to reproduce the reverb sound via the loudspeaker array, which is then mixed using the weight vector $\mathbf{WYXZR}^{(i)}$, and after energy compensation through Gammatone filters with gains $g_{rev}^{(i)}$, it is played back using conventional Ambisonics. Auditory transfer functions (ATFs) of BRIRs of two dummy heads in the recording and playback rooms are used as reference and target signals in the optimization procedure to find the gains of the Gammatone filters.

This allows the use of spherical harmonic representations for optimizing the rendering. To overcome common limitations of Ambisonics at high frequencies and to mitigate the detrimental effects of playback acoustics, the direct sound field components and the remaining diffuse sound field are first separated using beamforming techniques. This will allow a number of the limitations of conventional Ambisonics to be overcome. In Ambisonics-based sound reproduction, rendering high frequencies introduces a significant challenge due to the limited size of the sweet spot. Furthermore, the acoustic characteristics of the reproduction environment must be carefully compensated to ensure accurate spatial audio rendering. Inverse filtering methods in the Ambisonics domain have been proposed, as demonstrated in (Betlehem and Abhayapala, 2005) and (Lecomte *et al.*, 2018). However, significant challenges are still expected to remain at both low and high frequencies. Additionally, applying inverse filtering after conventional Ambisonics decoding introduces robustness issues. By capturing a separate direct sound field, this can be rendered using the VBAP method. In this way, the direction of arrival can be preserved, while at the same time the sound field at high frequencies will more closely represent the true direct sound field, including the interaural cues. Although ideally this direct sound field component would be the only component arriving at the ears of the listener, the beamforming will not be perfect, and additional reverberant components will be included in the beamformed signal as well. In addition, the playback via VBAP in the playback room will cause additional reverberant components. The central assumption in our approach is that these unintended reverberant components, as they appear at the listener's ears in the playback environment, will be lower in level than the reverberant components heard in the recording environment. As a consequence, to perceptually match the sound field heard in the recording environment, additional reverberation needs to be reproduced in the playback environment, specifically to compensate for what is missing when reproducing the VBAP components. A loudspeaker array is employed in the reproduction room to reproduce the reverberant sound, in principle using Ambisonics, but adapted to compensate for spectral and spatial distortions of the sound field caused by the playback environment. To achieve a perceptual match between the fully reproduced reverberant sound field and the reverberation in the recording venue, three perceptual criteria will be optimized. Firstly, the spectral shape of the fully reproduced reverberant sound field should match that of the recording venue. Secondly, the early decay curves of both the recording and reproduction rooms should be aligned. Thirdly, to preserve spatial impression, the interaural cross-correlation should also match between the recording and reproduction. Two dummy heads are placed in both the recording and playback rooms for the optimization process to enable comparison of these perceptual criteria. Although these perceptual criteria may initially seem limited in describing the full sound field, research in parametric spatial audio coding has shown that such perceptual descriptions can effectively represent spatial audio with high quality (Breebaart $et$ $al.$, 2005; Breebaart $et$ $al.$, 2007). To enable adjustment of the spectral properties of the direct and reverberant sound fields, as well as control over the energy decay curve of the combined sound field and the interaural cross-correlation, the spherical harmonic components are spectrally shaped using frequency dependent weighting factors ($g_{dir}^{(i)}$, $g_{rev}^{(i)}$ and $\mathbf{WYXZR}^{(i)}$) prior to rendering. These weighting factors are key parameters in the optimization process, which will be described in the following sections. Before the optimization process can be described, a precise description of the signal path from capture to rendering needs to be formulated. For this, in Section III, signal recording and reproduction is described. This section includes four subsections: sound field representation, direct and reverberant sound capturing, direct sound filtering and rendering, and finally, reverberant sound filtering and rendering.

## III. SIGNAL RECORDING AND REPRODUCTION

This section explains the signal representation in the Ambisonics domain, the beamforming process, and the separation of direct and reverberant sounds.

### A. Sound field representation

Here the precise definitions will be given such as used in this paper for representing spherical harmonics. Spherical harmonics (SHs) of order n and degree $m$ (Daniel, 2000) are defined as follows:

$$Y_n^m(\theta, \varphi) \equiv \sqrt{\frac{(2n+1)}{4\pi} \frac{(n-|m|)!}{(n+|m|)!}} P_{n,|m|}(\cos\theta) \begin{cases} \sqrt{2}\sin(|m|\varphi) & \text{if } m < 0 \\ 1 & \text{if } m = 0 \\ \sqrt{2}\cos(|m|\varphi) & \text{if } m > 0 \end{cases},$$

(1)

in which $\theta$ and $\varphi$ represent elevation and azimuth angles, respectively, and $P_{n,m}$ denotes the associated Legendre polynomial of order $n$ and degree $m$. Note that this definition does not include the Condon-Shortley phase (Arfken *et al.*, 2011).

The SHs expansion of a plane wave arriving from incidence angle $(\theta_s, \varphi_s)$ is given by:

$$e^{i\mathbf{kr}} = 4\pi \sum_{n=0}^{\infty} \sum_{m=-n}^{n} i^n j_n(kr) Y_n^m(\theta_s, \varphi_s) Y_n^m(\theta, \varphi),$$

(2)

where $i$ is the imaginary unit number, $k$ is the wavenumber, and $j_n(kr)$ is the $n^{\text{th}}$ order spherical Bessel function of the first kind. The SHs expansion for a point source at position $\mathbf{r}_s = (r_s, \theta_s, \varphi_s)$ with $r < r_s$ is given by:

$$\frac{e^{ik\|\mathbf{r}-\mathbf{r}_s\|}}{4\pi \|\mathbf{r}-\mathbf{r}_s\|} = ik \sum_{n=0}^{\infty} \sum_{m=-n}^{n} h_n^{(1)}(kr_s) j_n(kr) Y_n^m(\theta_s, \varphi_s) Y_n^m(\theta, \varphi),$$

(3)

where $h_n^{(1)}$ is the spherical Hankel function of the first kind and of order $n$. In practice, the infinite summation across $n$ in Eq.(3) is truncated to a finite order $N$. A general form of the SHs expansion of the recorded pressure by a microphone in a room is given by:

$$P(r,\theta,\varphi,k) = \sum_{n=0}^{N} \sum_{m=-n}^{n} A_n^m(k) b_n(kr) Y_n^m(\theta,\varphi), \qquad (4)$$

where $A_n^m(k)$ are the SHs coefficients, and $b_n(kr)$ is a parameter that depends on the scattering characteristics of the recording point. The recorded pressure $P(r,\theta,\varphi,k)$ is assumed to be the summation of the direct and reverberant sounds, which can be modeled by plane waves or point sources. The spatial information of both the direct and reverberant sounds is summed in the SHs coefficient $A_n^m(k)$. The factors $b_n(kr)$ represent the effect of having a rigid spherical microphone array with radius $r_e$, and are equal to:

$$b_n(kr) = j_n(kr) - \frac{j_n'(kr_e)}{h_n^{(2)'}(kr_e)} h_n^{(2)}(kr) \qquad (5)$$

The SHs coefficients are obtained from the recorded pressure of microphone capsules on a spherical microphone array:

$$A_n^m(k) = \frac{\sum_{q=1}^{Q} \alpha_q P(r_e,\theta_q,\varphi_q,k) Y_n^m(\theta_q,\varphi_q)}{b_n(kr_e)}, \qquad (6)$$

in which, $r_e$ is the radius of microphone array, $\theta_q$ and $\varphi_q$ are respectively the elevation and azimuth angles of microphone capsules on the rigid sphere, $Q$ is the number of microphone capsules on the array and $\alpha_q$ are element weights used to maintain the orthonormality of SHs in order to achieve limited discrete sampling of theoretically continuous spatial recordings (Rafaely, 2015).

### B. Direct and reverb sounds capturing

This study will only consider the capturing and rendering of a single source. Although, theoretically a single source, including room acoustics can be perfectly represented in the spherical domain, in practice, the maximum order of available Ambisonics recording microphones is limited to typically four (mh-acoustics, 2023) or six (mh-acoustics, 2025). Remaining within the low-order conventional Ambisonics domain can lead to significant issues, including spatial aliasing and poor low-frequency estimation during recording (Moreau *et al.*, 2006). At low frequencies, capturing high-order Ambisonics is challenging, as the Ambisonics signal often falls within the noise floor. Furthermore, the playback process using conventional Ambisonics is problematic due to its limited spatial resolution at low frequencies. Therefore, similar to DirAC **(Pulkki, 2007)**, in the approach presented here, the direct and reverberant sounds are considered separately. We use the EM-32 Eigenmike microphone from M-Acoustic (mh-acoustics, 2023) for sound recording, which supports 4th order Ambisonics ($N=4$). To capture the direct sound, an axis-symmetric beamformer with narrowest main beam (Rafaely, 2015) is directed towards the sound source, with a direction of arrival angle of $(\theta_{DOA}, \varphi_{DOA})$ :

$$S_{BF}(k) = \sum_{n=0}^{4} \sum_{m=-n}^{n} A_n^m(k) Y_n^m(\theta_{DOA}, \varphi_{DOA}). \qquad (7)$$

Here, the "*BF*" subscript stands for "beamformer". The elevation and azimuth of DOA $(\theta_{DOA}, \varphi_{DOA})$ can either be predetermined or estimated. In this paper only the predetermined DOA information is used for beamforming.

To capture the reverb signal, the direct sound is subtracted from the full recorded sound field. To accomplish this, the beamformed signal with the corresponding DOA (Eq. (7)) is spatialized using the plane-wave model from Eq.(2) and then subtracted from the encoded Ambisonics channels. The spatialized beamformed signal for order *n* and degree *m* is:

$$S_{n\,BF}^{m}(k) = S_{BF}(k)\Big[ Y_n^m(\theta_{DOA}, \varphi_{DOA}) \Big], \qquad (8)$$

and the reverberant field is obtained after subtraction from the recorded SHs coefficients:

$$S_{n\,\mathrm{Rev}}^{m}(k) = A_n^m(k) - S_{n\,BF}^{m}(k). \qquad (9)$$

Here, "*Rev*" stand for "Reverberation". In this computation, it is assumed, as mentioned earlier, that only one source is present.

### C. Direct sound filtering and rendering

The upper signal flow corresponds to the rendering of the direct sound. Ambisonics is avoided for direct sound rendering, as it can introduce artifacts at both low and high frequencies. Additionally, the direct sound field would need to be spatialized again using a plane-wave or point-source model, potentially leading to further inaccuracies. For these reasons, similar to the DirAC approach (Pulkki, 2007), VBAP (PULKKI, 1997) is used instead. In Ambisonics-based reproduction, all loudspeakers are involved, whereas in VBAP, only three loudspeakers are used. Based on the predetermined DOA of the source, three VBAP loudspeakers are selected, and their respective gains are calculated. These gains are represented in a VBAP gain vector:

$$\mathbf{G}_{vbap} = \begin{bmatrix} g_1 \\ g_2 \\ g_3 \end{bmatrix}, \qquad (10)$$

in which, $g_i$ represents the VBAP gain of loudspeakers number $i$. In the time domain, the signals from three VBAP loudspeakers are filtered using 4th order Gammatone filterbank (Hohmann, 2002) before playback. The filtering process uses frequency-dependent gains $g_{dir}^{(i)}$ in Gammatone filter number $i$. The impulse response of the Gammatone filter number $i$ is denoted by $\gamma^{(i)}[n]$. The gain $g_{dir}^{(i)}$ of Gammatone filter is applied to achieve spectral energy equalization of the direct sound within filter number $i$. The inverse Fourier transform of the recorded direct sounds $S_{BF}(k)$ from Eq.(7) is denoted as $s_{BF}[n] = F^{-1}\{S_{BF}(k)\}$. The final driving signals for the direct sound, using three VBAP loudspeakers in Gammatone band $i$, are represented in a driving vector as follows:

$$\mathbf{D}_{dir}^{(i)} = \mathbf{G}_{vbap}\, g_{dir}^{(i)}\, \gamma^{(i)}[n] * S_{BF}[n]. \tag{11}$$

Here, "$*$" is the convolution operator, and both $\mathbf{D}_{dir}^{(i)}$ matrix and $\mathbf{G}_{vbap}$ vector have three rows.

### D. Reverb sound filtering and rendering

After obtaining the direct sound driving signals for the loudspeakers, the driving signals related to the reverb sound are calculated. In conventional Ambisonics, the reverb signal $S_{n\,\text{Rev}}^{m}(k)$ can be reproduced in an anechoic room using a simple source approach (Poletti, 2005). The driving signal of loudspeaker number $l$ is:

$$D_{l,\text{Rev}}(k) = \frac{g_l(k)}{ikR^2} \sum_{n=0}^{4} \frac{e^{-ikR}}{h_n^{(1)}(kR)} \sum_{m=-n}^{n} S_{n\,\text{Rev}}^{m}(k)\, Y_n^m(\theta_l, \varphi_l), \tag{12}$$

In this formula, $R$ is the radius of reproduction array, $g_l$ is a frequency-depended and loudspeaker-dependent optional gain used to modify the reproduction pattern. This gain controls the main and side lobes that emerge from the truncation of the harmonics, such as the weights max-$r_E$ used in (Zotter and Frank, 2012). The max-$r_E$ weighting maximizes the energy towards the panning direction. In this study, a different type of weighting will be used for the Ambisonics channels. The recorded reverb Ambisonics signal, obtained from Eq.(9) is renamed for simplicity and placed within the matrix $\mathbf{S}_{\text{Rev}}$:

$$\mathbf{S}_{\mathrm{Re}\nu} = \begin{bmatrix} S^0_{0\,\mathrm{Re}\nu}(k) \\ S^{-1}_{1\,\mathrm{Re}\nu}(k) \\ S^0_{1\,\mathrm{Re}\nu}(k) \\ S^1_{1\,\mathrm{Re}\nu}(k) \\ \mathbf{S}^n_{m\,\mathrm{Re}\nu}(k) \end{bmatrix} \equiv \begin{bmatrix} W_{\mathrm{Re}\nu}(k) \\ Y_{\mathrm{Re}\nu}(k) \\ X_{\mathrm{Re}\nu}(k) \\ Z_{\mathrm{Re}\nu}(k) \\ \mathbf{R}_{\mathrm{Re}\nu}(k) \end{bmatrix}. \tag{13}$$

In this context, $W_{\mathrm{Re}\nu}(k)$ is related to the omni-directional $0^{\mathrm{th}}$ order Ambisonics , while three channels of $Y_{\mathrm{Re}\nu}(k)$, $X_{\mathrm{Re}\nu}(k)$ and $Z_{\mathrm{Re}\nu}(k)$ are related to the $1^{\mathrm{st}}$ order Ambisonics or B-Format, which have dipole patterns. The matrix $\mathbf{R}(k)$ has 21 rows, including all recorded Ambisonics channels for the remaining orders $2 \leq n \leq 4$. The Ambisonics coefficients of reverb sound are used to construct the driving signal of loudspeaker number $l$, where $1 \leq l \leq L$. According to the Eq.(12), the driving signal for each harmonic can be separated. For example, the driving signals for $W_{\mathrm{Re}\nu}(k)$ and $Y_{\mathrm{Re}\nu}(k)$, without considering $g_l(k)$ weights in Eq.(12) are:

$$\begin{aligned} W_l(k) &= \frac{1}{ik} \frac{e^{-ikR} W_{\mathrm{Re}\nu}(k)}{h_0^{(1)}(kR)} Y_0^0(\theta_l, \varphi_l) \\ Y_l(k) &= \frac{1}{ik} \frac{e^{-ikR} Y_{\mathrm{Re}\nu}(k)}{h_1^{(1)}(kR)} Y_1^{-1}(\theta_l, \varphi_l) \end{aligned} \tag{14}$$

In this way, all the driving signals in the frequency domain can be placed into another matrix, $\mathbf{D}_l$:

$$\mathbf{D}_l = \begin{bmatrix} W_l(k) \\ Y_l(k) \\ X_l(k) \\ Z_l(k) \\ \mathbf{R}_l(k) \end{bmatrix}. \tag{15}$$

The time-domain equivalent of $\mathbf{D}_l$ is $\mathbf{d}_l = F^{-1}\{\mathbf{D}_l\}$:

$$\mathbf{d}_l = \begin{bmatrix} w_l[n] \\ y_l[n] \\ x_l[n] \\ z_l[n] \\ \mathbf{r}_l[n] \end{bmatrix}. \tag{16}$$

Each row in Eq. (16) is filtered by a Gammatone filterbank. The filtered signal in band $i$ is obtained as follows:

$$\mathbf{d}_l^{(i)} = (\gamma^{(i)}[n] \otimes \mathbf{1}_{25\times1}) \overset{*}{\circledast} \mathbf{d}_l = \begin{bmatrix} w_l^{(i)}[n] \\ y_l^{(i)}[n] \\ x_l^{(i)}[n] \\ z_l^{(i)}[n] \\ \mathbf{r}_l^{(i)}[n] \end{bmatrix}. \tag{17}$$

Here, $\otimes$ represents the Kronecker product, and 25 is the number channels for 4th order Ambisonics. The vector $\mathbf{1}_{25\times1}$ is a $25\times1$ vector where all elements are equal to one. Therefore, $\gamma^{(i)}[n] \otimes \mathbf{1}_{25\times1}$ is a vector with 25 rows, where each row is equal to the filter $\gamma^{(i)}[n]$. The operator $(\overset{*}{\circledast})$ is defined as a row-wise matrix convolution, meaning that each row of the first matrix is convolved with the corresponding row in the second matrix. A mixing vector $\mathbf{WYXZR}^{(i)}$ is used to combine Ambisonics channels for the Gammatone band number $i$. In the next section, we will see how $\mathbf{WYXZR}^{(i)}$ is utilized to control the IC at the position of the listener in the playback room. The mixing vector $\mathbf{WYXZR}^{(i)}$ consists of the following elements:

$$\mathbf{WYXZR}^{(i)} = \begin{bmatrix} w^{(i)} \\ y^{(i)} \\ x^{(i)} \\ z^{(i)} \\ r^{(i)} \mathbf{1}_{21 \times 1} \end{bmatrix} \tag{18}$$

For the $0^{\text{th}}$ Ambisonics $W^{(i)}$ and the three dipole $2^{\text{nd}}$ order Ambisonics components $Y^{(i)}$, $X^{(i)}$ and $Z^{(i)}$, separate mixing coefficients $w^{(i)}$, $y^{(i)}$, $x^{(i)}$ and $z^{(i)}$ are used respectively. For the higher-order Ambisonics channels, denoted as $\mathbf{R}^{(i)}$ which correspond to orders of 2, 3 and 4, a single mixing coefficient $r^{(i)}$ is used for simplification. After mixing the Ambisonics channels, a Gammatone weighting factor $g_{rev}^{(i)}$ is applied. The final driving signal of each loudspeaker for the reverb sound is given by:

$$\mathbf{D}_{rev}^{(i)}[n] = g_{rev}^{(i)} \left( \mathbf{WYXZR}^{(i)} \right)^T \mathbf{d}_l^{(i)} \ , \ 1 \le l \le L. \tag{19}$$

The mixing vectors $\mathbf{WYXZR}^{(i)}$ combined with the reverb gains $g_{rev}^{(i)}$ control both IC and the energy of the reverb field in the reproduction room. Note that, unlike $g_l(k)$ in Eq.(12), frequency-dependent $\mathbf{WYXZR}^{(i)}$ is not loudspeaker-dependent and is applied identically to all loudspeakers.

The gains of the Gammatone filters for direct ($g_{dir}^{(i)}$) and reverb($g_{rev}^{(i)}$) sounds, and mixing coefficients of reverb channels ($\mathbf{WYXZR}^{(i)}$) are determined through an optimization procedure, as depicted in the optimization block in FIG.1. In this process, the reproduced binaural room impulse response (BRIR) in the playback room will be adapted in the optimization and compared with that of the reference dummy head in the recording room. The mathematical details of the optimization are further explained in the next section.

## IV.    OPTIMAZION

For perceptually-based room compensation, various types of information are crucial. In particular, for the perception of direct sound, the direction of arrival (DOA) is vital. This information is closely tied to spatial cues such as the interaural time difference (ITD) and interaural level difference (ILD) (Blauert, 1997), as well as the precedence effect (Litovsky *et al.*, 1999) , which emphasizes the importance of the first wavefront reaching the ears. This is achieved by reproducing the sound source from the desired DOA and applying frequency-dependent direct gains $g_{dir}^{(i)}$. Beyond direct sound, the reverberant field provides essential cues related to envelopment, diffusion, brightness, and coloration, all influenced by natural reverberation. Additionally, the ratio between direct and reverberant sound conveys valuable information about the perceived spaciousness and distance of the sound source (Mershon and King, 1975). To manage the reverberant sound field and balance the direct and reverberant energy components, reverberant gains $g_{rev}^{(i)}$ are applied. Another relevant important parameter is the IC between the left and right ears, which influences the perception of directionality, distance, and the spatial qualities of sound (Blauert, 1997). In our approach, the IC is controlled using a mixing coefficient, denoted as $\mathbf{WYXZR}^{(i)}$.

To determine $g_{dir}^{(i)}$, $g_{rev}^{(i)}$, and $\mathbf{WYXZR}^{(i)}$, two dummy heads are used in the recording and reproduction rooms. For this optimization, the BRIRs of the source (as a reference), the Room Impulse Responses (RIRs) of the microphones in the recording room, and the BRIRs of the playback loudspeakers in the reproduction room are required. The dummy head in the recording room is positioned near the recording microphone to capture spatial cues accurately.

### A.  Direct sound optimization

For the optimization of direct sound, the BRIR of the reproduced direct sound in the reproduction room, referred as the beamformed direct RIR, is required along with the entire reproduction chain, denoted as $BRinR_{BF}$ . The RIR of the recording microphone for the target source is beamformed , and the component $H_{BF}(k)$ and reverb component $H^m_{n\,\mathrm{Rev}}(k)$ are obtained following a similar approach as in Eq.(7) and Eq.(8). Here, instead of the recorded direct signal $S_{BF}(k)$ and the recorded reverb signal $S^m_{n\,\mathrm{Rev}}(k)$, we work with the RIRs of the direct and the reverb parts. The BRIR vector for the VBAP loudspeakers in the reproduction room at the left ear is given by:

$$\mathbf{BRIR}_{dir,left} = \begin{bmatrix} BRIR_{SP_1,left}[n] \\ BRIR_{SP_2,left}[n] \\ BRIR_{SP_3,left}[n] \end{bmatrix} \tag{20}$$

For simplicity in notation, we omit the *left* subscript from this point forward. However, it is important to note that the $\mathbf{BRIR}_{dir}$ and other BRIRs can still be considered for both the left and right ears. According to Eq.(11), the $BRinR_{BF}$ is obtained by convolving the BRIRs of three loudspeakers (SP$_1$, SP$_2$, SP$_3$) with the beamformed direct RIR in time domain $h_{BF}(n)$, while incorporating all processing steps:

$$BRinR_{BF}[n] = g^{(i)}_{dir}\,\gamma^{(i)}[n] * h_{BF}[n] * \left( \mathbf{G}^T_{vbap} \mathbf{BRIR}_{dir} \right) \tag{21}$$

The reference BRIR recorded in the recording room can be decomposed into its direct and reverberant components:

$$BRIR_{ref}[n] = BRIR_{ref,dir}[n] + BRIR_{ref,rev}[n]. \tag{22}$$

The $BRinR_{BF}$ in Eq.(21) can also be decomposed into its direct and reverb components:

$$BRinR_{BF}[n] = BRinR_{BF,dir}[n] + BRinR_{BF,rev}[n]. \tag{23}$$

It was shown in (Grosse and van de Par, 2015) that the separation times between $BRIR_{ref,dir}[n]$ and $BRIR_{ref,rev}[n]$, as well as the separation time between $BRinR_{BF,dir}[n]$ and $BRinR_{BF,rev}[n]$, play an important role in determining the reverberation time ($T_{30}$). In this study, these parameters have been set empirically rather than optimized. Instead of attempting to make, $BRIR_{ref,dir}$ and $BRinR_{BF,dir}[n]$ identical, as is done in conventional inverse filtering, we compensate for their respective spectral energies. This approach ensures a more robust perceptually driven filtering process. It is implemented using a Gammatone analysis and synthesis framework (Hohmann, 2002), which enables adjustment to the "Auditory Transfer Function" (ATF) of the measured reference and reproduced BRIRs. The $ATF_{sig}^{(i)}$ is defined as the energy of a signal at the output of filter number $i$ of Gammatone filterbank. The goal of direct sound filtering is to equalize the spectral energies of the beamformed direct signal, reproduced by VBAP in the reproduction room (Eq.(22)), with the spectral energy of direct part of reference signal $BRIR_{ref,dir}$ (Eq(23)):

$$ATF^{(i)}\left(BRIR_{ref,dir,left}[n]\right) = \left(g_{dir,left}^{(i)}\right)^2 ATF^{(i)}\left(BRinR_{BF,left}[n]\right). \tag{24}$$

We added the subscript "left" to indicate that different gain values can be obtained for the left and right ears. For gain optimization, we employ a fast-converging method proposed in (Van De Par *et al.*, 2005), which ensures that negative gains are avoided for overlapping Gammatone filters. This optimization approach was also used in (Grosse and van de Par, 2015). A weighted average of the gains obtained for the left and right ears can be applied for the loudspeakers, ensuring balanced and perceptually accurate reproduction:

$$g_{dir}^{(i)} = x\, g_{dir,left}^{(i)} + (1-x)\, g_{dir,right}^{(i)}, \quad 0 \le x \le 1. \tag{25}$$

The x value can be set to 0.5 for normal averaging. The reconstructed direct sound after direct-sound optimization using $g_{dir}^{(i)}$ for Gammatone Filterbank synthesis is:

$$BRinR_{BF,opt}[n] = \underset{i=1:N_G}{\text{GammatoneSynthesis}}\left\{g_{dir}^{(i)} BRinR_{BF}^{(i)}[n]\right\} \tag{26}$$

Here, $N_G$ represents the number of Gammatone filters and the $BRinR_{BF,opt}[n]$ denotes the optimized direct signals. However, similar to Eq.(23), the $BRinR_{BF,opt}[n]$ can once again be decomposed into its direct and reverb components:

$$BRinR_{BF,opt}[n] = BRinR_{BF,opt,dir}[n] + BRinR_{BF,opt,rev}[n]. \qquad (27)$$

Here, the $BRinR_{BF,opt,dir}[n]$ represents the optimized direct sound, while the $BRinR_{BF,opt,rev}[n]$ refers to an unavoidable component of the direct sound caused by the reverberation of the playback room. This component is taken into account when optimization the reverb sound in the following section.

## B. Reverb sound optimization

A parameter required for optimization is the reproduced binaural RinR impulse response of the reverberant room, denoted as $BRinR_{Rev}$. This impulse response is obtained by convolving the recorded reverberant RIRs with the BRIRs of all loudspeakers in the reproduction room, taking into account the entire processing chain. To obtain the reverberant Room Impulse Response (RIR) in the SH domain, the spatialized direct signal is subtracted from the original recorded RIR in the SH domain, similarly to Eq. (9). A notation similar to that used in Eqs.(13)-(19) can be applied, substituting the signals with the BRIRs and RIRs. Following the approach presented in Eq.(13), a recorded reverberant RIR in the Ambisonics domain can be expressed as:

$$\mathbf{H}_{Rev} = \begin{bmatrix} H_{0Rev}^0(k) \\ H_{1Rev}^{-1}(k) \\ H_{1Rev}^0(k) \\ H_{1Rev}^1(k) \\ \mathbf{H}_{mRev}^n(k) \end{bmatrix} \equiv \begin{bmatrix} W_h(k) \\ Y_h(k) \\ X_h(k) \\ Z_h(k) \\ \mathbf{R}_h(k) \end{bmatrix}. \qquad (28)$$

Here, the subscript "$h$" indicates that the Ambisonic signals correspond to impulse responses. Similarly to Eqs.(16) and (17), the driving signal for loudspeaker number $l$, this time for the recorded reverberant RIRs, can be expressed at the output of the Gammatone filter number $i$ as:

$$\mathbf{d}_{h,l}^{(i)} = \begin{bmatrix} w_{h,l}^{(i)}[n] \\ y_{h,l}^{(i)}[n] \\ x_{h,l}^{(i)}[n] \\ z_{h,l}^{(i)}[n] \\ \mathbf{r}_{h,l}^{(i)}[n] \end{bmatrix}. \tag{29}$$

By considering Eq.(19) together with the BRIRs of loudspeaker array $(BRIR_l, 1 \le l \le L)$, $BRinR_{\mathrm{Rev}}$ cab be obtained as follows:

$$BRinR_{\mathrm{Rev}}^{(i)}[n] = g_{rev}^{(i)} \left( \mathbf{WYXZR}^{(i)} \right)^T \left( \sum_{l=1}^{L} \mathbf{d}_{h,l}^{(i)} \circledast \left( BRIR_l \otimes \mathbf{1}_{25 \times 1} \right) \right). \tag{30}$$

Here, the driving signals of loudspeaker number $l$ comprising all Ambisonics channels filtered by $i$-th Gammatone filter, denoted as $\mathbf{d}_{h,l}^{(i)}$, are convolved with the respected $BRIR_l$. After summing across all loudspeakers in the array, those signals are weighted using loudspeaker-independent mixing coefficients $\left( \mathbf{WYXZR}^{(i)} \right)$. Subsequently, the energy-compensating gain $g_{rev}^{(i)}$ is applied for each channel. It should again be noted that both $BRIR_l$ and $BRinR_{\mathrm{Rev}}^{(i)}[n]$ can be considered for either the left or right ears. For optimizing the reverberant RIR, the IC between the left and right ears is taken into account. The main idea to compensate the IC in the center of the loudspeaker array involves using only two Ambisonics channels: the W and Y channels (Fallah and van De Par, 2020). The W channel, representing the 0th-order Ambisonics, exhibits an omnidirectional pattern, while the Y channel, representing one of the 1st-order components, features a dipole pattern aligned with the line connecting the left and right ears of the dummy head in both the recording room and the center of the reproduction array. The W channel is intended to produce an in-phase signal, whereas the Y channel generates an out-of-phase signal between the ears. Consequently, a weighted combination of these two channels is expected to establish the desired interaural correlation (IC). By using only the W and Y channels, the $BRinR_{\text{Rev}}$ in Eq.(30) can be modified to:

$$BRinR_{\text{Rev},WY}^{(i)}[n] = g_{rev}^{(i)} \left( \mathbf{WY}^{(i)} \right)^T \left( \sum_{l=1}^{L} \begin{bmatrix} w_{h,l}^{(i)}[n] \\ y_{h,l}^{(i)}[n] \end{bmatrix} \circledast \left( BRIR_l \begin{bmatrix} 1 \\ 1 \end{bmatrix} \right) \right). \tag{31}$$

In this approach, it is also possible to select specific combinations of Ambisonics channels for example $BRinR_{\text{Rev},WY}$ , $BRinR_{\text{Rev},WYX}$ , $BRinR_{\text{Rev},WYZ}$ or $BRinR_{\text{Rev},XYXZ}$ instead of utilizing all Ambisonic channels in $\mathbf{WYXZR}^{(i)}$. Here, the W and Y channels are always retained due to their essential role in controlling IC. For these type of truncated Ambisonic representations, the corresponding mixing vectors are denoted as $\mathbf{WY}^{(i)}$, $\mathbf{WYX}^{(i)}$, $\mathbf{WYXZ}^{(i)}$ and $\mathbf{WYZ}^{(i)}$, respectively. However, in the general case, all Ambisonics channels, as indicated in Eq.(30), are utilized for the reproduction of the reverberant field. Nonetheless, the impact of using a truncated set of Ambisonics channels can be investigated. It has been shown that, for Ambisonics-based headphone reproduction, preserving higher-order components is beneficial (Engel *et al.*, 2021). In our approach, adding the X channel to the basic WY configuration is expected to enhance robustness during reproduction, particularly when listeners change their head position or orientation. Another important consideration is the complexity of optimization and computational load. To reduce this complexity, separate mixing weights are applied only to the $0^{\text{th}}$ and $1^{\text{st}}$-order Ambisonics channels (W, Y, X, and Z), while a constant weight is assigned to the remaining higher-order channels, as defined in Eq.(18). In the general case where all Ambisonic channels are used as described in Eq.(30), the total reproduced reverberant BRIR in the playback room is obtained by summing the reverberant components from Eq.(27) and Eq.(30):

$$BRinR^{(i)}_{Total\,\mathrm{Re}v}[n] = BRinR^{(i)}_{BF,opt,rev}[n]$$
$$+g^{(i)}_{rev}\left(\mathbf{WYXZR}^{(i)}\right)^{T}\left(\sum_{l=1}^{L}\mathbf{d}^{(i)}_{h,l}\,\overline{*}\left(BRIR_{l}\otimes\mathbf{1}_{4\times1}\right)\right). \tag{32}$$

In Eq(32), the signal $BRinR_{BF,opt,rev}[n]$ is included because the VBAP loudspeakers used for reproducing the direct sound are generate a non-separable reverberant component. The combination of Ambisonics mixing coefficients $\mathbf{WYXZR}^{(i)}$ and the gain factors $g^{(i)}_{rev}$ enables control over the reproduced IC and the overall reverberant energy, respectively. In cases where the $T_{30}$ of the reproduction room is shorter than that of the recording room, the $BRinR_{BF,opt,rev}[n]$ in Eq.(27) contains less reverberation energy than required, potentially resulting in an under-represented or less immersive reverberant field. Therefore, additional reverberation can be added according to Eq.(32). A general observation about this approach is that it only supports scenarios where the reproduction room has a $T_{30}$ that is equal to or shorter than that of the recording room. Fortunately, this condition typically holds true in practice. Recordings are often made during concert events held in large halls, while reproduction usually takes place in smaller rooms. To determine the optimal values of the $\mathbf{WYXZR}^{(i)}$ vector for controlling IC between left and right ear signals, a grid search algorithm is employed. This algorithm explores various combinations of parameters $w^{(i)}$, $y^{(i)}$, $x^{(i)}$, $z^{(i)}$ and $\mathbf{r}^{(i)}$, using predefined value ranges with a constant step size for each:

$$-[\alpha\,\beta\,\chi\,\delta\,\varepsilon]^{T} \leq [w^{(i)}\,y^{(i)}\,x^{(i)}\,z^{(i)}\,r^{(i)}]^{T} \leq [\alpha\,\beta\,\chi\,\delta\,\varepsilon]^{T}. \tag{33}$$

In this way, for each combination of $[\alpha\,\beta\,\chi\,\delta\,\varepsilon]$, similar to the direct sound energy optimization in described in Eq.(24), an optimal gain $g^{(i)}_{rev}$ can be determined. This gain ensures that the ATF of the reproduced reverb field matches that of the reverberant reference:

$$ATF^{(i)}\left(BRIR_{ref,rev,left}[n]\right) =$$
$$ATF^{(i)}\left(BRinR_{BF,opt,rev,left}^{(i)}[n]\right) +$$
$$\left(g_{rev,left}^{(i)}\right)^2 ATF^{(i)}\left(\left(\mathbf{WYXZR}^{(i)}\right)^T\left(\sum_{l=1}^{L}\mathbf{d}_{h,l}^{(i)}\,\overline{*}\left(BRIR_l \otimes \mathbf{1}_{25\times1}\right)\right)\right). \tag{34}$$

Here, following the assumption in (Grosse and van de Par, 2015), the two reverberant components, $BRinR_{BF,opt,rev}[n]$ and $BRinR_{\mathrm{Re}v}^{(i)}[n]$, are considered uncorrelated. Similar to the gain averaging approach used for VBAP loudspeakers in Eq.(25), a weighted average of the left and right ear gains can be used as the final gain for the loudspeaker array in each frequency band:

$$g_{rev}^{(i)} = y\,g_{rev,left}^{(i)} + (1-y)g_{rev,right}^{(i)} \quad 0 \le y \le 1. \tag{35}$$

This allows for balanced energy adjustment across both ears. For playback of the compensated reverberant sound, all loudspeakers in the array are utilized. Both the $\mathbf{WYXZR}^{(i)}$ and $g_{rev}^{(i)}$ are applied to filter the reverberant Ambisonics signal of each loudspeaker prior to playback. The total reproduced BRIR in each frequency band is obtained by summing the optimized direct component from Eq.(27) and the reverberant component, which is processed using a predefined matrix $\mathbf{WYXZR}^{(i)}$ and its corresponding optimized $g_{rev}^{(i)}$ from Eq.(30), as follows:

$$BRinR^{(i)}[n] = BRinR_{BF,opt}[n] + BRinR_{\mathrm{Re}v}^{(i)}[n] \tag{36}$$

For this signal, the interaural cross correlation (IACC) and IC for Gammatone band number ($i$) are calculated as follows:

$$IACC_{BRinR^{(i)}}[q] = \frac{\sum_{m=-\infty}^{m=\infty} BRinR_{left}^{(i)}[m]\ BRinR_{right}^{(i)}[m+q]}{\sqrt{\sum_{m=-\infty}^{m=\infty}\left(BRinR_{left}^{(i)}[m]\right)^2}\ \sqrt{\sum_{m=-\infty}^{m=\infty}\left(BRinR_{right}^{(i)}[m]\right)^2}} \tag{37}$$

$$IC_{BRinR^{(i)}} = \max(IACC_{BRinR^{(i)}})$$

The IC between $BRinR_{left}^{(i)}$ and $BRinR_{right}^{(i)}$, denoted as $IC_{BRinR}^{(i)}$ is computed and compared with the reference IC, $IC_{ref}^{(i)}$. The reference IC is calculated between the signals $BRIR_{ref,left}^{(i)}[n]$ and $BRIR_{ref,right}^{(i)}[n]$ as defined in Eq.(22). For each combination of mixing coefficients $[\alpha\ \beta\ \chi\ \delta\ \varepsilon]^T$ and its corresponding gain $g_{rev}^{(i)}$, an absolute error function is calculated. The combination that minimizes the IC error is selected as the final reverb-compensating mixing coefficients and gain for each frequency band:

$$\min_{\mathbf{WYXZR}^{(i)}, g_{rev}^{(i)}} \left\{ IC_{error}(i) \right\} = \min_{\mathbf{WYXZR}^{(i)}, g_{rev}^{(i)}} \left\{ \left\| IC_{BRinR_{opt}}^{(i)} - IC_{ref}^{(i)} \right\| \right\}. \tag{38}$$

The optimization ensures that the reproduced reverberant signal closely matches the spatial characteristics of the reference in terms of IC. The optimization procedure for both direct and reverb components is summarized in Algorithm.(1). As mentioned earlier, it is possible to use only first order Ambisonics (B-Format) and even truncated versions of 1$^{st}$ order Ambisonics, instead of the full $\mathbf{WYXZR}^{(i)}$. To investigate the impact of using different number of Ambisonics channels, several configurations are considered in the optimization process, including the full B-format $\mathbf{WYXZ}^{(i)}$, and the truncated versions $\mathbf{WY}^{(i)}$, $\mathbf{WYX}^{(i)}$, and $\mathbf{WYZ}^{(i)}$. For instance, in the case of $\mathbf{WY}^{(i)}$, only the coefficients $w^{(i)}$ and $y^{(i)}$ are optimized. The grid search algorithm is then limited to searching within the defined ranges, $-\alpha \le w^{(i)} \le \alpha$ and $-\beta \le y^{(i)} \le \beta$. This approach allows for analysis of how reducing the number of Ambisonics channels effects the accuracy of spatial reproduction and interaural coherence control.

Algorithm.1. Calculation of the optimal parameters $g_{dir}^{(i)}$, $\mathbf{WYXZR}^{(i)}$ and $g_{rev}^{(i)}$.

**Step 1**: Determination of direct sound energy-compensating gains $g_{dir}^{(i)}$ using Eq.(24) and Eq.(25).

**Step 2:** Determination of optimal reverb mixing coefficients $\mathbf{WYXZR}^{(i)}$ and energy-compensating gains $g_{rev}^{(i)}$:

**(1)** Setting the elements of $\mathbf{WYXZR}^{(i)} : [w^{(i)} \, y^{(i)} \, x^{(i)} \, z^{(i)} \, r^{(i)}]^T$ to a predefined values according to Eq.(33).

**(2)** An optimal $g_{rev}^{(i)}$ is determined for the set of $[w^{(i)} \, y^{(i)} \, x^{(i)} \, z^{(i)} \, r^{(i)}]^T$ in **(1)** according to the Eq.(34) and Eq.(35).

**(3)** Calculation of the total reproduced RIR $BRinR^{(i)}[n]$ in Eq.(36).

**(4)** The reproduced IC between left and right ears for $BRinR^{(i)}[n]$ and is calculated according to Eq.(37).

**(5)** The error function for ICs of reference and reproduced BRIRs are calculated:
$$IC_{error}(i) = \left| IC_{BRinR}(i) - IC_{ref}(i) \right|.$$

**Step 3:** A grid searching for $-\alpha \le w^{(i)} \le \alpha$ , $-\beta \le y^{(i)} \le \beta$ , $-\chi \le x^{(i)} \le \chi$ , $-\delta \le z^{(i)} \le \delta$ and $-\alpha \le r^{(i)} \le \alpha$ with the defined step size and repeating the sub-steps (1), (2),(3), (4) and (5) in **Step 2** to select a different $\mathbf{WYXZR}^{(i)}$ and its corresponding $g_{rev}^{(i)}$ that give the minimum error in Eq.(38).

## V. EVALUATION OF THE METHOD

### A. Measurements

Two sets of recordings in different rooms were conducted to evaluate the proposed method. The first set was recorded at the Theater Laboratorium, and the second at St. Christophorus Church, both located in Oldenburg, as depicted in FIG.2. The $T_{30}$ value in the theater is approximately *500 ms*, whereas in the church it is around *1500 ms*.

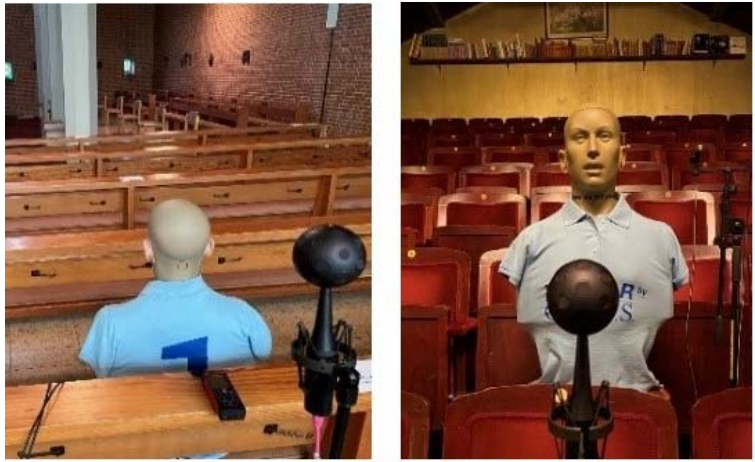

FIG.2. Measurement setups were deployed in two acoustically different rooms in Oldenburg. Right: Theater Laboratorium ($T_{30} \approx 500\ ms$). Left: St. Christophorus Church ($T_{30} \approx 1500\ ms$). In both rooms, a KEMAR dummy head microphone and an Eigenmike were used to capture BRIRs and spatial RIRs, respectively.

Measurements were conducted to compare acoustic characteristics under different reverberation conditions. The mh-acoustics EM32 Eigenmike (mh-acoustics, 2023) and the KEMAR dummy head ((n.d.)) were employed as recording microphones, while Dynaudio Core 5 loudspeakers were used to play the excitation sweeps for measuring RIRs and BRIRs. The Eigenmike was employed to capture both room impulse responses (RIRs) and recordings of music and speech. Two microphone positions were selected in each recording room. In the first position (Pos. 1), the microphone was placed almost directly in front of the source at an approximate distance of 10 meters. In the second position (Pos. 2), the source was located to the left of the microphone at an approximate distance of 15 meters and an azimuthal angle of about 30 degrees. The positions of the KEMAR dummy head and the Eigenmike were arranged to minimize mutual acoustic interference from reflections. To achieve this, the recording devices were placed at slightly different heights. In both positions in the theater and in Position 1 of the church, the Eigenmike was placed slightly lower and in front of the KEMAR dummy head. However, in the church at Position 2, the dummy head was placed on a bench; therefore, the Eigenmike was positioned behind the dummy head and slightly higher. In addition to the RIR and BRIR measurements, live recordings of voice, piano, and violin performances were conducted in the Theater Laboratorium, along with a voice recording in the church. All recordings were made using an RME Fireface UFX+ audio interface at a sampling frequency of 44.1 kHz.

For audio reproduction, a loudspeaker array based on a 50-node Lebedev grid (Lecomte *et al.*, 2016) was used, enabling the encoding and playback of 5th-order Ambisonics signals through real loudspeakers. This setup is located in the loudspeaker laboratory of the Acoustics Research Group at the University of Oldenburg, as depicted in FIG.3. The array employs compact Genelec 8010 loudspeakers and has a radius of approximately 1.5 meters. The average T30 measured at the center of the array in this reproduction environment is approximately 500 ms.

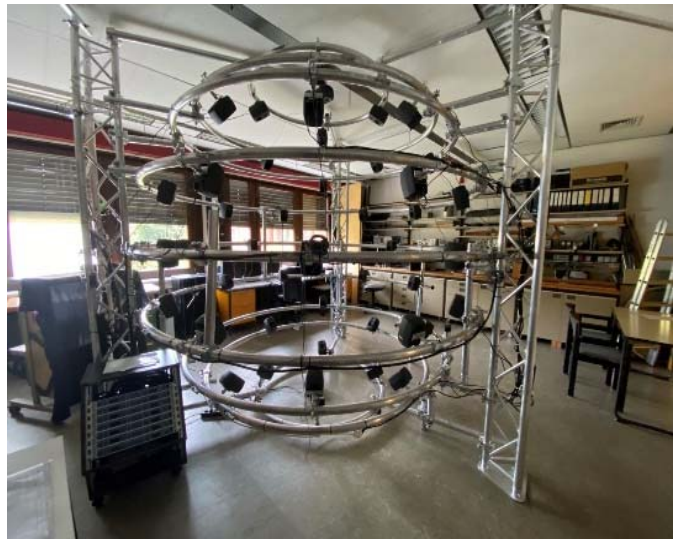

FIG.3. Loudspeaker array setup in the loudspeaker laboratory of the Acoustics Group in Oldenburg, featuring 50 loudspeakers arranged according to a 50-node Lebedev grid. The array has a radius of approximately 1.5 meters and is located in an environment with an average $T_{30}$ of about *500 ms* at the central listening position. This setup was used for the spatial reproduction of 5th-order Ambisonics signals.

## B. Optimization results

For the optimization, the recorded BRIRs and RIRs captured respectively with the KEMAR dummy head and the Eigenmike in the recording rooms as well as the BRIRs of 50 loudspeakers measured with KEMAR in the reproduction room, were used. At a sampling frequency of *44.1 kHz*, 42 Gammatone filters were employed for analysis and synthesis in the proposed approach. The optimization results for the processed RIRs, compared to the reference BRIRs in the theater at Position 2, are shown in FIG.4. In the optimization process, as defined in Eq.(22), the first *10 ms* of the reference BRIRs were considered the direct part, while the remainder was treated as the reverberant part. The theater environment, with a $T_{30}$ of approximately *500 ms* that is comparable to that of the playback room, presents a challenging scenario for preserving spatial and spectral characteristics during reproduction. The two upper panels in FIG.4 display the Acoustic Transfer Functions (ATFs) of the reproduced BRIRs for the left and right ears. These panels compare the ATFs of four processed versions (WY, WYX, WYXZ, and WYXZR) with those of the reference (Ref) and the conventional unprocessed Ambisonics case (UnP). It is important to note that in the UnP case, the recorded Ambisonics signal is decoded using Eq.(12), where the reverberant component $S_{n \text{Rev}}^{m}(k)$ is replaced by the complete recorded Ambisonics signal $A_{n}^{m}(k)$, and then reproduced without any further processing.

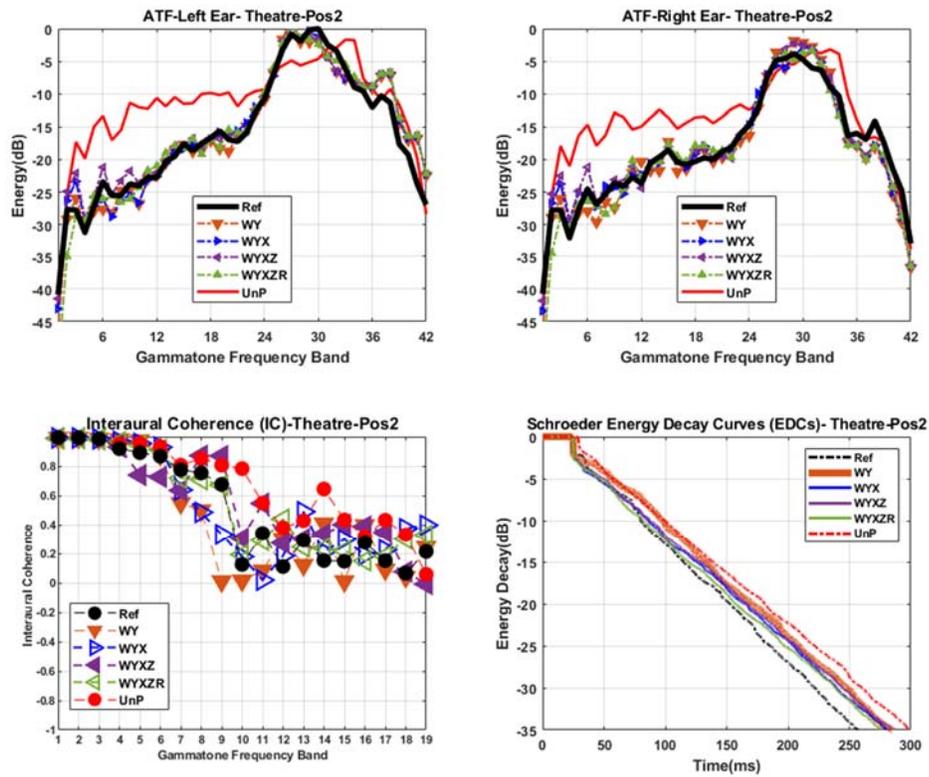

FIG.4. Comparison of four processed rendering cases (WY, WYX, WYXZ, and WYXZR) with the reference (Ref) and the conventional Ambisonics case (UnP) in the theater. The two upper panels show the acoustic transfer functions (ATFs) for the left and right ears across the reference, UnP, and processed cases. The processed methods exhibit improved spectral energy alignment with the reference across most Gammatone frequency bins. The lower-left panel illustrates the interaural cues (ICs) of the processed and UnP cases compared to the reference for perceptually relevant frequencies below *1500 Hz*. Improvements are observed across most frequency bands in all processed cases, except for bins 5–11, where only the WYXZR case shows a strong match. The lower-right panel presents Schroeder energy decay curves (EDCs) of the BRIRs, demonstrating enhanced decay characteristics and reverberation time alignment in the processed cases relative to the UnP case.

As expected, due to the higher amount of reverberation, the ATFs of the conventional Ambisonics case (UnP) exhibit noticeable deviations from the reference, particularly at low frequencies where reverberant energy is more dominant. In contrast, the ATFs of the processed versions more closely align with those of the reference across most Gammatone frequency bands. However, in some low-frequency bands where reverberation strongly dominates (e.g., bands 3 and 4), perfect compensation is not fully achieved. The lower-left panel of FIG.4 compares the interaural coherence (IC) values of the processed and conventional Ambisonics BRIRs with those of the reference signal across frequency bands up to band 19, corresponding to frequencies below 1500 Hz. This range is perceptually critical, as the auditory system is particularly sensitive to changes in interaural coherence (IC) at these frequencies. For Gammatone bands between 5 and 11, notable differences are observed among the compensation methods. The WYXZR case demonstrates the best alignment with the reference in this range, whereas the WY and WYXZ (B-Format) cases show little to no improvement. There is a fundamental limitation due to the way spherical harmonics are affected by the reproduction room acoustics. The playback room tends to make the reproduced sound field more diffuse. On the one hand this likely brings the reproduced sound field nearer to the recorded sound field in terms of IC. On the other hand, there is no mathematical guarantee that the precise intended interaural coherence (IC) pattern can be fully restored. In practice, adjusting the strength of individual harmonics helps to compensate to a good degree, but the achievable match will not be perfect. Most Gammatone filterbank outputs for the compensated cases show improved IC, outperforming the conventional Ambisonics case (UnP), which lacks room compensation.

The lower-right panel of FIG.4 presents the Schroeder energy decay curves (EDCs) (Schroeder, 1965) for the reference, unprocessed, and processed cases. These curves show that the proposed method leads to improved $T_{30}$ values. Among the processed cases, the EDCs for WYXZ and WYXZR more closely align with the reference compared to those of WYX and WY. This improvement is primarily attributed to the limited low- and mid-frequency content in the higher-order Ambisonics channels captured by the Eigenmike recordings (mh-acoustics, 2023). Consequently, the WYXZ and WYXZR cases contribute less low-frequency reverberant energy during the optimization process, resulting in a shorter reverberation tail that better matches the reference.

The results for the church environment at Position 2 are shown in FIG.5. The church exhibits a significantly higher $T_{30}$ (~*1500 ms*) compared to the playback room (~*500 ms*), which facilitates the preservation of spatial and spectral characteristics during reproduction. According to the upper panels of FIG.5 and similar to the results obtained for the theater, the ATFs of the processed cases achieve a much closer match to the reference across a wide range of Gammatone frequency bins. For interaural coherence (IC), with the exception of a few Gammatone filters in certain cases, the processed signals show significant improvement over the conventional Ambisonics case, demonstrating better alignment with the reference. These improvements in IC suggest a more accurate spatial impression and enhanced localization cues during playback. For the EDCs, the highly reverberant nature of the church environment compared to the playback room provides greater flexibility for the proposed compensation method. The processed cases exhibit a closer alignment with the reference decay curve, indicating effective reverberation compensation. The improved matching observed in the WYXZR case is partly attributed to the reduced contribution of low-frequency reverberation in the higher-order Ambisonics channels, as captured by the Eigenmike. This reduction leads to less added reverberation in these channels, resulting in a shorter and more accurate reverberation tail that more closely resembles the original acoustic environment. The main difference between the theater and the church arises from their reverberation times. When the recording room has a longer $T_{30}$ than the playback room as in the case of the church, there is sufficient acoustic space for the recorded reverberation to be added to the direct sound.

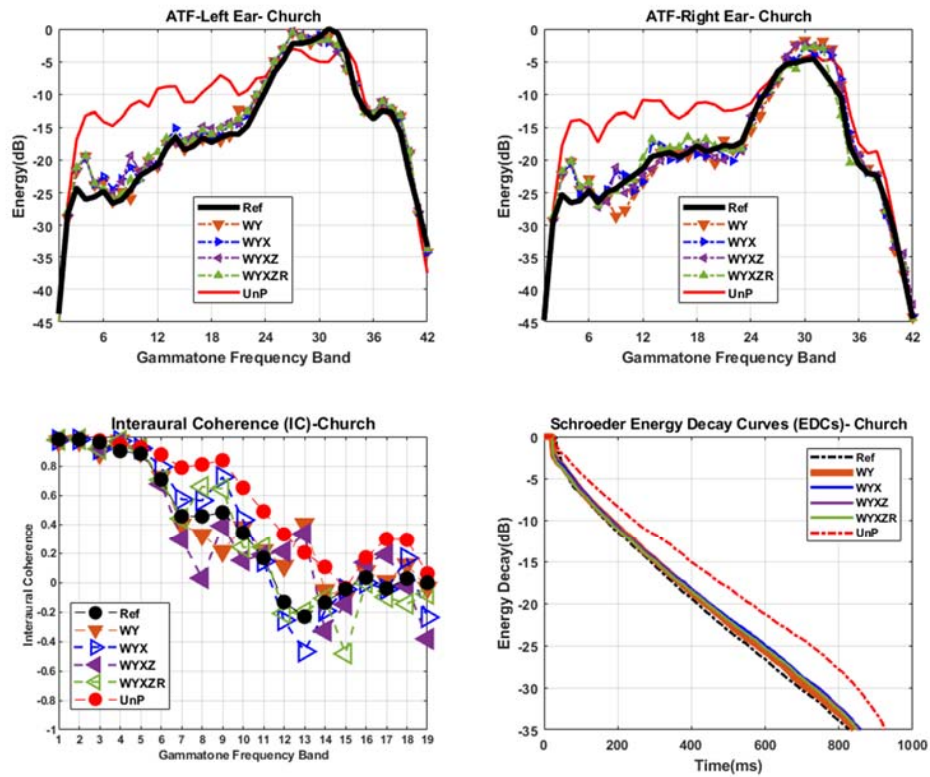

FIG.5. Similar to FIG.4, but showing results for the church environment. The compensated cases exhibit improved energy alignment across most Gammatone frequency bins. The processed cases also demonstrate significantly better interaural coherence (IC) alignment with the reference compared to the conventional Ambisonics approach, except for a few frequency bins in some cases. For the EDCs, all processed cases closely replicate the reverberation characteristics of the reference, with shorter and more accurate decay profiles compared to the UnP case. These results highlight the effectiveness of the proposed method in typical scenarios where the $T_{30}$ of the recording room is longer than that of the playback room.

This condition allows greater flexibility in the compensation process across the entire frequency range. In contrast, when the recording and playback rooms have similar $T30$ values, as in the theater, our approach remains largely effective by compensating for frequency regions where coloration has been altered by the playback room. Overall, the results from both environments confirm the effectiveness of the proposed method in reproducing the spectral and spatial characteristics of the recordings, even when the recording and playback rooms exhibit similar reverberation times.

### C. Subjective evaluation

For the subjective evaluation of the proposed method, a multi-stimulus listening test based on the MUSHRA framework (BS, 2003) was conducted for two headphone experiments and three loudspeaker rendering experiments. The purpose was to find how closely the proposed method matches the sound field recorded at the recording sites and how the proposed method compares to existing methods. Fourteen participants (6 male, 8 female) with normal hearing, aged between 21 and 35 years (mean age: 26), took part in both the headphone and loudspeaker experiments. The conditions presented to participants included: (i) a reference signal recorded with a dummy head in the recording room (Ref), (ii) a low-pass filtered version of the reference used as an anchor (Anker), (iii) a signal reproduced in a reverberant room using conventional 4th-order Ambisonics with 25 channels and no room compensation (ConvAmb), (iv) an ideal 4th-order Ambisonics rendering simulated under ideal conditions using 50 loudspeakers (IdAmb), (v) an optimized direct sound only rendered using Vector Base Amplitude Panning (VBAP), and (vi) five versions of the compensated signal generated using the proposed method, labeled WY, WYX, WYXZ, WYXZR, and WYZ. Note that the IdAmb condition was included only in the first headphone-based experiment. In total, 10 conditions were evaluated in the first headphone experiment, and 9 conditions were evaluated in both the second headphone experiment and the loudspeaker experiment. Listeners were instructed to assign higher scores to audio samples that more closely resembled the reference signal, either in terms of overall similarity, or similarity in terms of a specific evaluation criterion. In the headphone experiments, only the general similarity to the reference was assessed, whereas in the loudspeaker experiment, overall quality, timbral naturalness, and spatial naturalness were evaluated all in terms of similarity to the presented reference. The test stimuli were generated by convolving five dry (anechoic) audio signals: snare drum, guitar, clarinet, and a male voice, with the corresponding BRIRs. Additionally, real recordings were captured using the Eigenmike array: a male voice, a piano piece, and a violin piece in the theater, and a male voice in the church. For both the theater and church conditions, only Position 2 was used for the convolved stimuli. For the real recordings in the theater, the piano performance corresponds approximately to Position 2, while the voice and violin recordings were captured near Position 1. In the church, the real recording (male voice) was captured at Position 1.

### 1. Headphone Evaluations

This experiment was conducted with 14 listeners using headphones in a listening booth. Audio signals were presented at 70 dB SPL through Sennheiser HD 650 headphones, driven by an RME UFX+ soundcard, and equalized using AKtools (Brinkmann and Weinzierl, 2017) to ensure an almost flat frequency response. The headphone evaluation consisted of two parts. In the first part, the final rendering was simulated using recorded BRIRs from 50 loudspeakers, similar to those used in the optimization process. This setup allows for evaluating the method's effectiveness when the rendering and optimization are performed at the same positions for the convolved stimuli. However, for real voice, violin, and piano recordings captured in the theater, as well as voice recordings from the church, mismatches still exist between the recording and optimization positions. For these real recordings, the evaluation offers insight into the method's robustness with respect to variations on the recording side. For the first part of the headphone experiment, the mean scores and standard errors are shown in FIG.6. The upper two panels illustrate the variability across subjects for each instrument recorded in the theater (upper-left) and the church (upper-right).

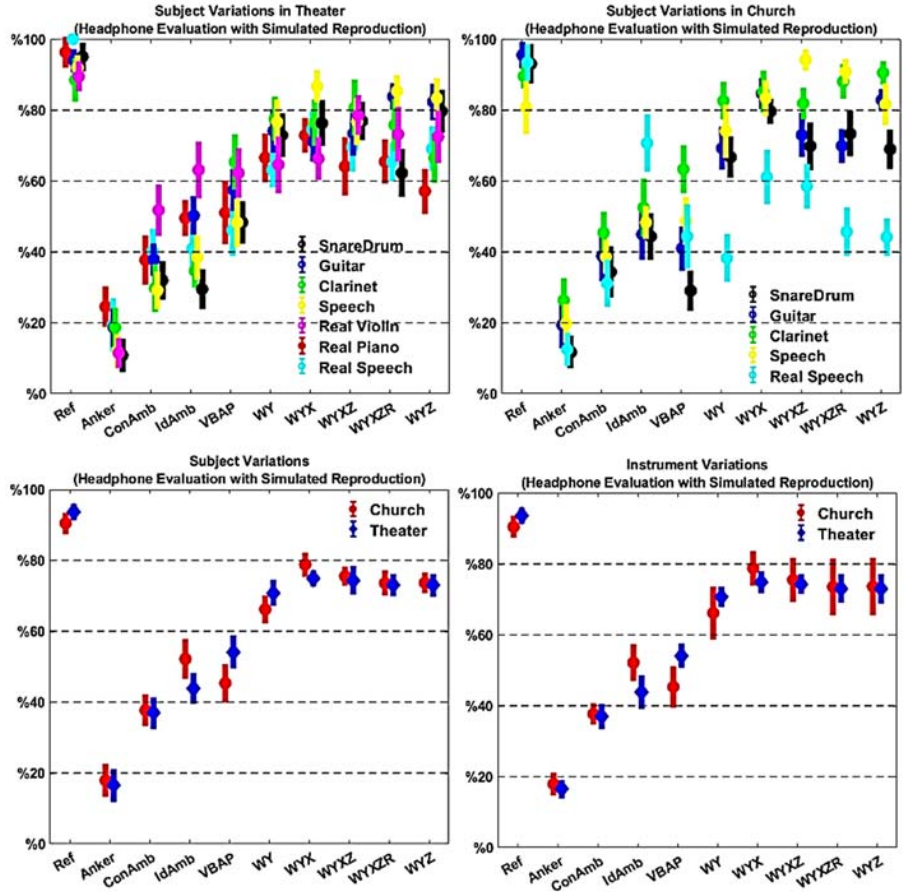

FIG.6. General quality assessment results from the first part of the headphone experiment. The mean scores and standard errors across subjects are shown for various audio signals in the theater (upper-left) and church (upper-right). The lower panels display the overall mean and standard errors across all subjects and instruments. The lower-left panel shows variation across subjects, while the lower-right panel illustrates variation across instruments. The participants' ratings indicate that conventional Ambisonics (ConvAmb) and ideal Ambisonics (IdAmb) are consistently rated lower than the compensated cases (WY, WYX, WYXZ, WYXZR, WYZ) across all instruments, except for the noisy speech recording in the church. Additionally, the results show that the compensated direct signal (VBAP) achieves notably better-quality ratings than ConvAmb.

In both panels, the hidden references (Ref) are not always correctly identified by the listeners, indicating that some test conditions produced signals of similar perceptual quality to the reference. In both environments, the Anker signals were almost always identified, reflecting their consistently low perceived quality. In the upper-left panel (theater recordings), the average scores for all instruments using conventional Ambisonics (ConvAmb) are lower than those of the compensated cases: WY, WYX, WYXZ, WYXZR, and WYZ. A similar trend is observed for the ideal Ambisonics (IdAmb), which involves the ideal reproduction of 4th-order Ambisonics. This suggests that 4th-order Ambisonics recording and reproduction by itself is insufficient, and compensation is necessary, even when the reproduction room does not include reflections. This limitation is primarily due to the small sweet spot for frequencies above 3 kHz in 4th-order Ambisonics reproduction, which is even smaller than the average human head size (Stitt *et al.*, 2014). Interestingly, in the theater, a relatively dry environment with a high direct-to-reverberant ratio (DDR), the mean scores for ideal Ambisonics (IdAmb), derived from the ideal reproduction of 4th-order Ambisonics, are consistently lower than those of the compensated direct sound (VBAP) for all instruments except the real violin. This highlights the importance of applying compensation, even when it is applied only to the direct sound. In the church condition, shown in the upper-left panel, a similar trend is observed for all audio signals except for the real speech. The SNR for the real speech recording is low due to the distant placement of the talker from the Eigenmike and the presence of background noise. Additionally, the microphone capsules of the Eigenmike generate a certain amount of self-noise, which becomes more noticeable when the recorded signal level is low. This issue does not occur during the recording of room impulse responses (RIRs), as the loudspeaker levels can be set to higher values. It is important to mention that compensating noisy Eigenmike recordings can result in colored noise, which tends to be more annoying than the original recorded noise. To reduce this effect for the speech recordings made in the church, the average noise profile across all stimuli was added to each stimulus, including the reference.

Despite this compensation, for the real speech recording in the church, only the WYX and WYXZ cases exhibit higher mean scores compared to the VBAP and ConvAmb cases. In contrast to the theater, in the church, only the clarinet receives higher scores with VBAP than with IdAmb; in all other cases, IdAmb is rated higher. This can be explained by the fact that the church is a more reverberant space, where accurate rendering of the reverberant signal becomes more critical to perceived audio quality.

In the two lower panels of FIG.6, the averaged results for subject variation (i.e., averaging first over instruments) and instrument variation (i.e., averaging first over subjects) are shown in the lower-left and lower-right panels, respectively. In both rooms, the use of conventional Ambisonics (ConvAmb) results in significant signal degradation, with average scores around 40% on a 100% scale. Ideal Ambisonics (IdAmb) achieves slightly higher scores, ranging between 40–60%. This level of degradation suggests that even ideal 4th-order Ambisonics, when rendered in an anechoic environment, fails to deliver perceptual quality comparable to that of a dummy head recording. The compensated direct sound (VBAP) provides only directional cues and does not fully reproduce all the desired characteristics of the reference signal. Its scores range between 40–60%, yet still remain higher than those of conventional Ambisonics (CoAmb). In the theater, a dry room with low reverberation, VBAP scores are higher than those of ideal Ambisonics (IdAmb). In contrast, in the church, where reverberation is stronger, VBAP scores fall below those of IdAmb. In the compensated WY case, which utilizes only two Ambisonics channels, the required transmission data is significantly reduced compared to conventional higher-order Ambisonics. Nevertheless, the mean scores for the WY case consistently exceed those of CoAmb, ranging between 60–70% in both environments. The results indicate that incorporating additional Ambisonics channels during reproduction can enhance audio quality. Notably, even the inclusion of just one extra channel, as seen in the WYX and WYZ cases, leads to noticeable improvement. In particular, the WYX case achieves mean scores approaching 80%.

For the subject variations shown in the lower-left panel of FIG.6, the statistical difference between the WYX and WY cases in the theater is not significant ($t(138) = 1.44$, $p = 0.15$), whereas in the church, the difference is significant ($t(138) = 3.33$, $p < 0.01$). For the WYZ and WY cases, the ratings do not show significant differences in either the theater ($t(138) = 0.70$, $p = 0.48$) or the church ($t(138) = 1.88$, $p = 0.062$). The WYXZ case, also referred to as the B-format, does not show a significant improvement compared to the WYX case in either the theater ($t(138) = -0.20$, $p = 0.84$) or the church ($t(138) = -0.96$, $p = 0.33$). Similarly, the compensated 4th-order Ambisonics case (WYXZR) does not show significant improvement over the WYXZ case in the theater ($t(138) = -0.40$, $p = 0.68$) or the church ($t(138) = -0.51$, $p = 0.61$). These results suggest that, within our approach, using just the WY channels along with the X (or possibly Z) channel is sufficient to achieve high-quality audio reproduction.

In the second part of the headphone evaluation, instead of using simulated rendering, the filtered audio files were physically played through a 50-loudspeaker array and recorded using a dummy head positioned at the center of the array. This experiment is important for two main reasons. First, it involves real reproduction, allowing for a more accurate comparison by capturing the complete chain of recording, processing, and playback without any simulation. Second, it enables an assessment of the method's robustness under realistic conditions. The robustness issue relates to two factors: (1) the displacement of the dummy head during real rendering compared to its original position during the dummy head recordings, and (2) changes in the room environment between the two measurement sessions, which were conducted a few months apart. The results of this experiment are presented in FIG. 7. Since the second headphone experiment does not involve simulation, the ideal Ambisonics (IdAmb) condition is excluded from this part of the evaluation.

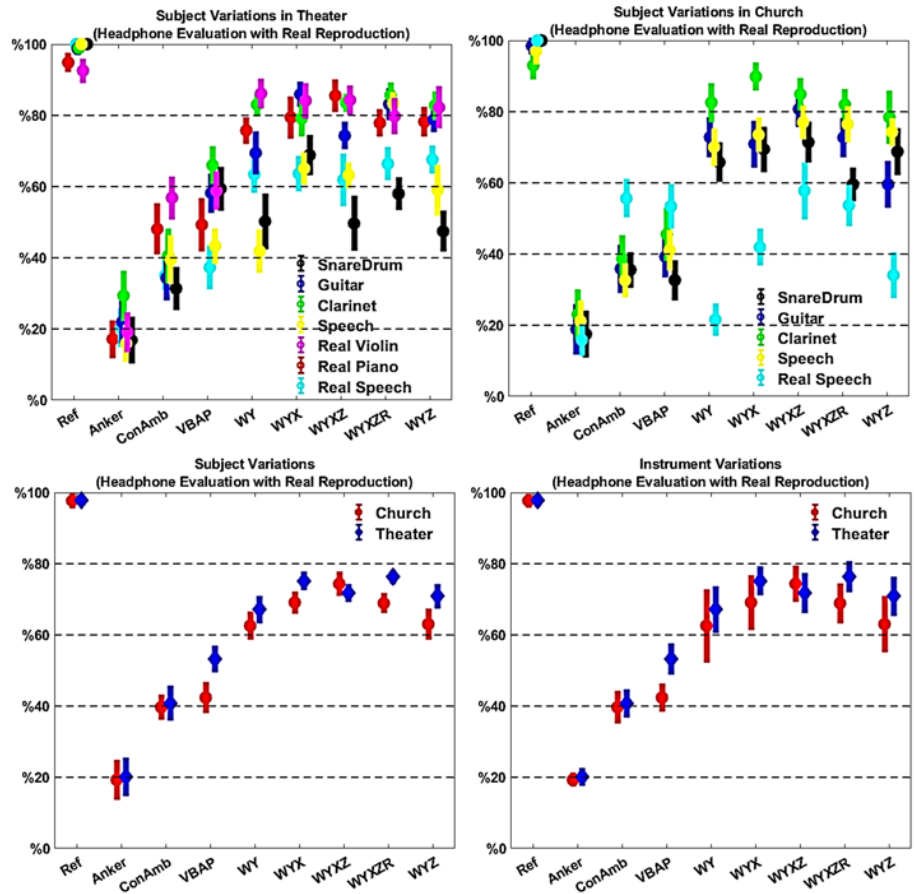

FIG. 7. Results from the second part of the headphone experiment using real audio playback in the loudspeaker lab. This experiment evaluates the robustness of the method, taking into account factors such as dummy head displacement and changes in the room environment over several months. While the overall trend is similar to that observed in the first headphone experiment FIG.6, some differences emerge due to ambient noise, particularly for the snare drum in the theater, where only the WYX case among the compensated methods outperforms VBAP. In the church, the lower scores for real speech reproduction reflect the impact of colored noise introduced during the processing stage, with only the WYXZ and WYXZR cases achieving scores comparable to VBAP and conventional Ambisonics (ConvAmb). Increasing the number of channels helps reduce the effect of this noise, with the WYXZR case showing the most noticeable improvement.

For individual instruments, the overall scoring trends are similar to those observed in the first part of the evaluation. All compensated cases including WY, WYX, WYXZ, WYXZR, and WYZ, as well as the VBAP case, receive higher scores than the conventional Ambisonics (ConvAmb) case. The main difference between the first and second headphone experiments lies in the ambient noise level. In the second experiment, the ambient noise is audible at normal signal levels, which was not the case in the first part of the headphone experiment. Since the BRIR recordings of the loudspeaker array used in the first headphone experiment were conducted with a high SNR, ambient noise had a negligible impact on the quality of the main signal. There are two noticeable differences compared to the first headphone experiment. First, for the snare drum in the theater, only the WYX case among the compensated cases achieved a higher mean score than VBAP. The snare drum, as an impulsive instrument with a broader frequency spectrum than the others, may reveal limitations in robustness across the full frequency range and may also be affected by the presence of ambient noise. The second major difference concerns the reproduction of real speech in the church. In the second headphone experiment, the scores for the compensated cases were significantly lower. In this context, only the WYXZ and WYXZR cases achieved scores comparable to those of the VBAP and ConvAmb cases. This suggests that the colored noise introduced during the processing stage had a greater impact on perceived quality than the benefits offered by room compensation. As the number of reproduction channels increases, the influence of noise is progressively reduced. This reduction in noise is due to the spatial summation effect at the center of the array, particularly evident in the compensated WYXZR and ConvAmb cases, where all Ambisonics channels are utilized in the rendering process. These findings indicate that under conditions where the signal level is weak, such as during speech in large reverberant spaces like churches, the impact of microphone and ambient noise becomes more pronounced, which can reduce the effectiveness of this approach. In such cases, the highest perceived quality is achieved with the VBAP method, which effectively attenuates recording noise through beamforming. For both subject and instrument variations shown in the two lower panels, the quality improvements in the compensated cases are statistically significant compared to the ConvAmb case. Statistical analysis was conducted for instrument variations to compare the compensated cases. The difference between the WYX and WY cases in the theater is significant ($t(138) = 2.20, p < 0.05$), but not in the church ($t(138) = 1.65, p = 0.11$). The differences between the WYZ and WY cases are not significant in either the theater ($t(138) = 0.37, p = 0.65$) or the church ($t(138) = 0.09, p = 0.92$). The WYXZ case does not show a significant improvement in quality compared to the WYX case in either the theater ($t(138) = -1.36, p = 0.18$) or the church ($t(138) = 1.48, p = 0.14$). Similarly, no significant difference is observed between the WYXZR and WYXZ cases in the theater ($t(138) = 0.62, p = 0.50$) or the church ($t(138) = -0.07, p = 0.94$). This analysis confirms the findings from the first headphone evaluation and suggests that, within our approach, the WYX configuration, comprising the omnidirectional (W) and two horizontal dipole (X and Y) channels, is sufficient to achieve the highest possible perceptual quality.

### 2. Loudspeaker Evaluation

For these experiments, participants were seated at the center of the loudspeaker array and asked to rate the stimuli played through the loudspeakers. They were also allowed to listen to the reference audio through headphones; however, no hidden reference was included among the test conditions. Participants were encouraged to slightly move or rotate their heads during the experiment, allowing the robustness of the methods to be evaluated under natural listening conditions. The loudspeaker evaluation was conducted across three sessions: general quality assessment in the first session, timbral assessment in the second, and spatial naturalness in the third. The results of the general quality assessment are shown in FIG. 8. The general trend of improvement for the compensated cases is consistent with the results from the headphone evaluation. However, there are some notable differences. In the headphone experiment, particularly in its first part, the scores for direct VBAP rendering were generally higher than those for ConvAmb. In contrast, for the loudspeaker experiment, the scores for VBAP and ConvAmb were, on average, very close to each other in both the theater ($t(138) = 1.05$, $p = 0.29$) and the church ($t(138) = 0.61$, $p = 0.53$).

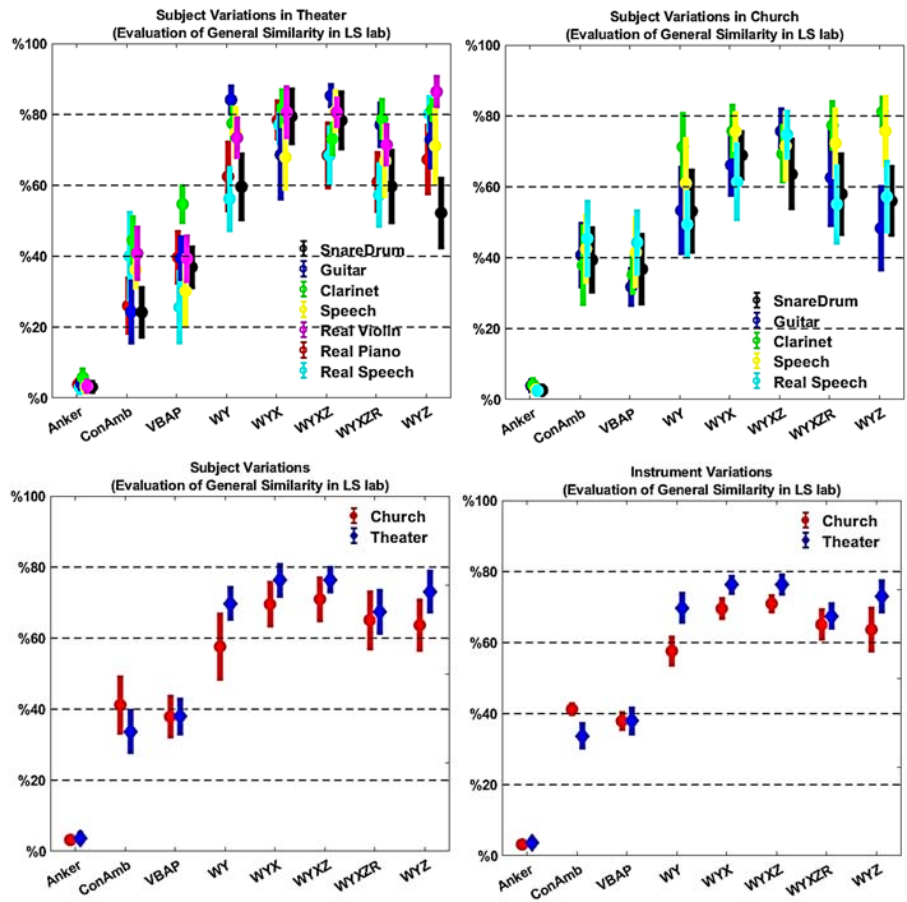

FIG. 8. Comparison of quality in the loudspeaker evaluations. While the overall trends are similar to those observed in the headphone evaluation, some differences are evident. In particular, the noisy recording in the church receives higher scores compared to the headphone evaluation. Overall, the WYX and WYXZ cases achieve the highest scores. Adding higher-order channels (WYXZR) reduces quality in the theater, likely due to limitations in Ambisonics reproduction. These results suggest that three Ambisonics channels (WYX) are sufficient for effective spatial audio rendering.

This suggests that the headphone evaluation does not fully capture certain aspects of real loudspeaker reproduction. Another interesting finding relates to the scoring of the noisy speech recording in the church. Unlike in the headphone evaluation, participants gave higher scores to the compensated signals, particularly in the WYX and WYXZ cases. This indicates that the presence of ambient noise in the recordings had a greater impact during the headphone evaluation compared to the loudspeaker evaluation. In both rooms, the highest mean scores were observed for the WYX and WYXZ cases. This is particularly evident in the theater for the snare drum, which has the broadest frequency spectrum among the tested instruments. For subject variations, the difference between the WYX and WY cases in the theater ($t(138) = 1.66$, $p = 0.09$) is not statistically significant, whereas in the church ($t(138) = 2.06$, $p < 0.05$) it is significant. Furthermore, the differences between the WYZ and WY cases are not significant in either room ($t(138) = 0.82$, $p = 0.41$ in the theater; $t(138) = 0.95$, $p = 0.34$ in the church). The same holds true when comparing the WYXZ and WYX cases, with no significant difference observed in the theater ($t(138) = 0.05$, $p = 0.95$) or the church ($t(138) = 0.29$, $p = 0.76$). Interestingly, however, the difference between the WYXZR and WYXZ cases is statistically significant in the theater ($t(26) = -2.41$, $p < 0.05$). In this case, adding all Ambisonics channels to the B-format results in a decrease in quality. This occurs because higher-order Ambisonics components (orders 2 to 4) are not well controlled, at low frequencies due to high noise levels, and at high frequencies due to the very small size of the sweet spot. However, in the church, no significant difference was observed between the WYXZR and WYXZ cases ($t(26) = -1.03$, $p = 0.30$). This suggests that using only three Ambisonics channels, one omnidirectional (W) and two horizontal dipoles (X and Y), is sufficient for rendering reverberant sound. Including all Ambisonics channels does not lead to improved quality in the rendered audio.

The second loudspeaker experiment focuses on spectral or timbral similarity. According to ITU-R BS.2399-0 (BS, 2017), the term that more precisely describes this evaluation is "Full," which refers to the overall timbral impression encompassing both low- and high-frequency content. As in the first loudspeaker experiment, participants were allowed to listen to the reference audio via headphones and were permitted to move and rotate their heads during this session. The results of the timbre experiment are shown in FIG. 9.

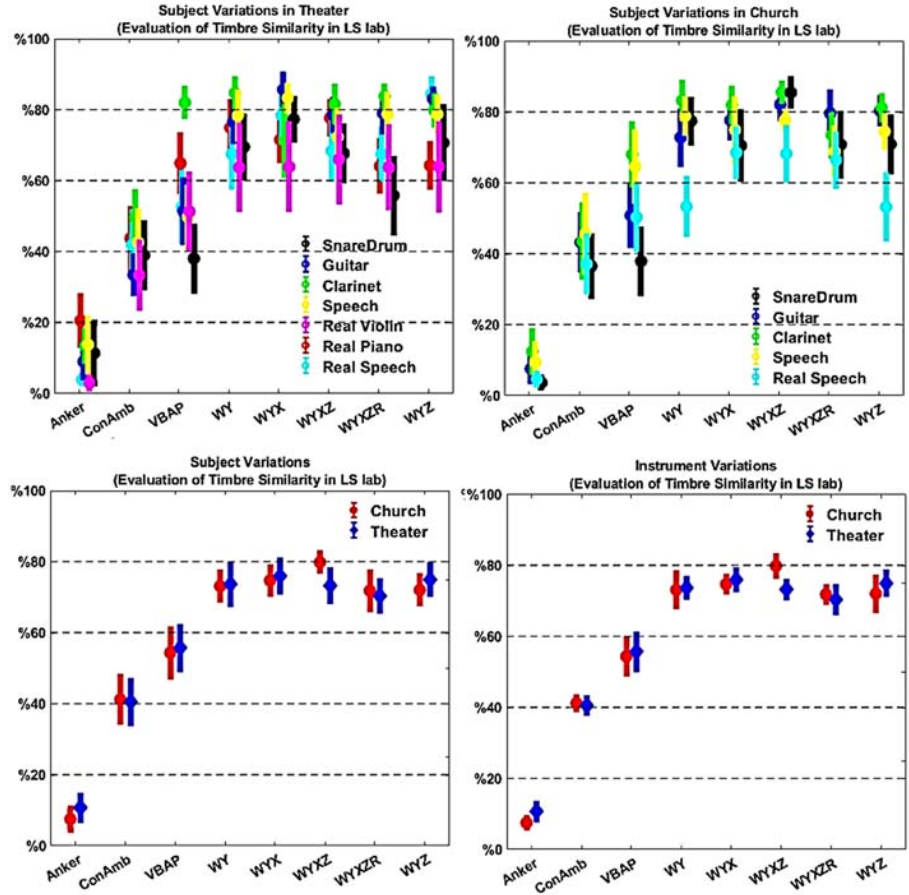

FIG. 9. Results of the timbral comparison evaluation in both the theater and church environments. The compensated VBAP method yielded higher mean scores than conventional Ambisonics (ConvAmb). On average, the WYX case performed best in the theater, while the WYXZ case showed the highest performance in the church. The findings suggest that adding additional Ambisonics channels does not significantly enhance timbral quality and may, in some cases, lead to a reduction in overall perceived quality.

When participants were asked to evaluate based on timbre, the compensated VBAP method received significantly higher mean scores than the conventional Ambisonics (ConvAmb) case. This difference was statistically significant in both the theater ($t(138) = 3.29$, $p < 0.01$) and the church ($t(138) = 2.31$, $p < 0.05$). This contrasts with the general quality evaluation, where listeners gave nearly identical scores to VBAP and ConvAmb. Based on the subject and instrument variation data presented in the lower panels, the WYX case in the theater and the WYXZ case in the church achieved the highest mean scores. However, the differences between WYXZ and WYX cases were not statistically significant in either the theater ($t(138) = -0.69$, $p = 0.49$) or the church ($t(138) = 1.36$, $p = 0.17$). Similarly, no significant differences were found between WYX and WY cases in the theater ($t(138) = 0.56$, $p = 0.57$) or the church ($t(138) = 0.35$, $p = 0.72$). These results indicate that adding more Ambisonics channels does not substantially affect timbral perception. Furthermore, similar to the general loudspeaker evaluation, the inclusion of all Ambisonics channels does not lead to an improvement in quality and may, in some cases, result in a decline. The difference between WYXZR and WYXZ cases in the theater was not significant ($t(138) = -0.70$, $p = 0.48$), but in the church, it was significant ($t(138) = -2.16$, $p < 0.05$). In summary, the results of the timbre evaluation, like those of the general quality experiment, suggest that increasing the number of Ambisonics channels does not lead to a noticeable improvement in perceived audio quality.

In the third session of the loudspeaker experiment, participants assessed the spatial naturalness of the stimuli. They evaluated how naturally spacious each stimulus sounded and how effectively it conveyed the sense of presence within the original acoustic environment, assigning higher scores to stimuli exhibiting superior performance in these perceptual attributes. As in previous sessions, participants were allowed to freely move and rotate their heads during the evaluation. The results obtained from the spatial naturalness evaluation are depicted in FIG. 10.

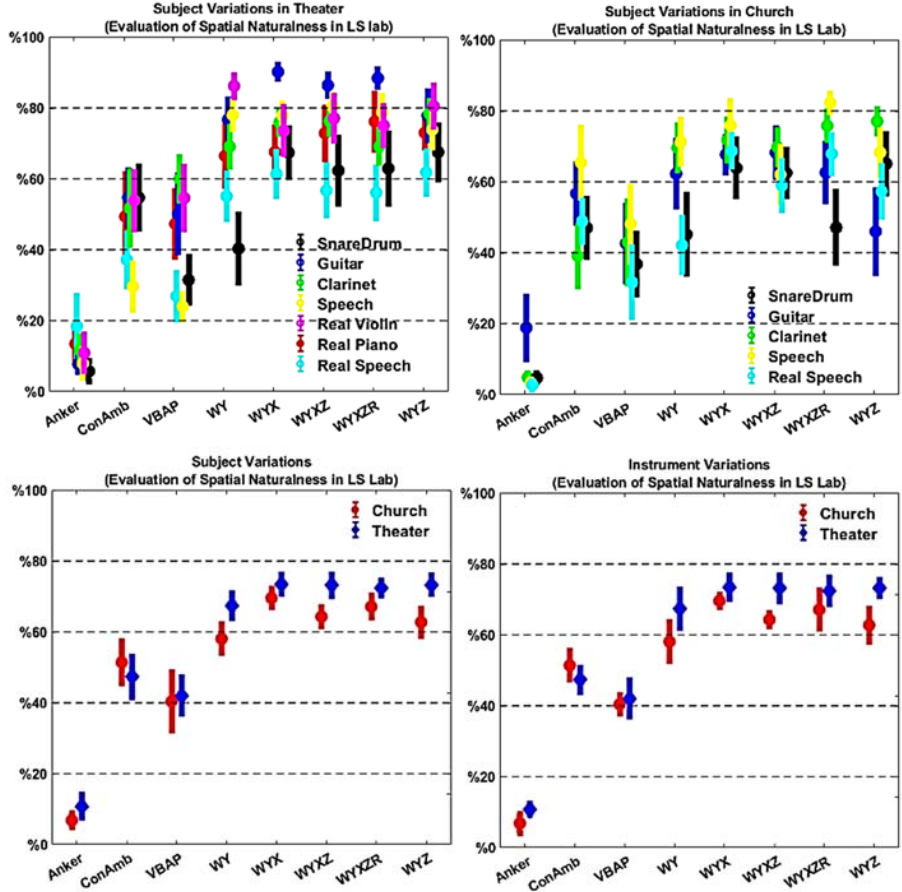

FIG. 10. Results of the spatial naturalness evaluation for the theater and church environments. Participants assessed stimuli based on how effectively each condition conveyed the impression of being present in the original room while freely moving and rotating their heads. The evaluation also examined the robustness of the reproduction methods. Unlike the timbral evaluation, conventional Ambisonics (ConAmb) scored higher than compensated VBAP in terms of spatial naturalness, mainly due to the perceived dryness of VBAP reproduction. In the theater, no significant differences were found among the compensated conditions. In the church environment, the WYXZR case exhibited higher spatial naturalness scores than other compensated cases, except for the WYX case, which achieved the highest scores overall.

The main difference compared to the timbral evaluation in FIG. 9 relates to the scores of the VBAP and ConAmb conditions. In the lower panels of FIG. 10, the mean scores of ConAmb exceed those of the compensated VBAP. This result is generally expected since participants can readily perceive that signals reproduced using only VBAP sound drier compared to other stimuli. The difference between VBAP and ConAmb in the church environment is statistically significant ($t(138)$ = $-1.95$, $p < 0.05$), whereas in the theater, this difference is not significant ($t(138) = -1.12$, $p = 0.26$). This finding can be explained by the lower reverberation time in the theater, which reduces the influence of reverberant characteristics on perceived reproduction quality. For the theater, no significant differences are found among the other conditions regarding spatial naturalness. Interestingly, the WYXZR case in the theater shows similar mean scores to the WYX, WYXZ, and WYZ cases when evaluating spatial naturalness. In the church environment, the difference between the WYX and WY cases is statistically significant ($t(138) = 2.39$, $p < 0.05$). Furthermore, the WYZ case generally achieves lower mean scores compared to the WYX case. This outcome is attributed to participants primarily performing head movements in the horizontal rather than vertical plane, making the presence of the 'X' pattern critical for robustness in horizontal localization. Adding additional Ambisonics channels doesn't notably enhance the perceived quality; for instance, the difference between WYX and WYXZ is not significant ($t(138) = -0.07$, $p = 0.93$). These results confirm, as in previous experiments, that the WYX condition offers an optimal compromise between perceived quality and the number of channels required.

## VI.    DISCUSSION

In this study, a perceptually motivated recording and reproduction approach based on Ambisonics was proposed. This approach not only compensates for the acoustics of the playback room but also implicitly considers the filtering required by conventional 4th order Ambisonics for high-quality rendering. Rather than aiming for an ideal 'physical' reconstruction of the original sound field, the proposed method accurately reproduces the directional cues, energy distribution, spatial diffuseness, and reverberation characteristics of the recorded audio. To render the direct sound, the Vector Base Amplitude Panning (VBAP) method was employed, utilizing beamforming techniques. The reverberant sound field was reproduced through various Ambisonics configurations: full 4th-order Ambisonics (WYXZR, 25 channels), complete 1st order Ambisonics (WYXZ), and truncated 1st-order Ambisonics versions (WY, WYX, WYZ). Both the direct and reverberant signals were preprocessed using frequency-dependent mixing coefficients and energy compensation gains implemented within a Gammatone-based analysis and synthesis framework. These parameters were obtained via an optimization procedure utilizing RIRs and BRIRs measured in both the recording and playback rooms. The optimization aimed to match certain perceptual characteristics of the reproduced BRIRs at the listener's position in the playback room to those directly measured in the original recording environment.

By reproducing the direct sound using the VBAP approach, most directional cues are preserved. The compensation procedure separately addresses energy for both the direct and reverberant sounds, as well as the interaural coherence (IC) for their combined signals. Through this energy-based compensation, the reverberation time characteristics of the original recordings are implicitly maintained

Both room impulse responses (RIRs) and real recordings captured with an Eigenmike microphone array in a theater and a church were used for the optimization and evaluation of the proposed approach. For reproduction, a loudspeaker array arranged according to a Lebedev grid was placed in a reverberant room.

The results of all headphone and loudspeaker listening experiments indicated that, for every version of the compensated signals, the general quality, timbre, and spatial characteristics of the reproduced sound were significantly improved compared to conventional Ambisonics without compensation. This improvement is also observed when the compensated reproductions are compared with ideally rendered Eigenmike recordings in a simulated anechoic environment. These findings highlight that raw Ambisonics recordings alone are insufficient for high-quality audio reproduction, and that equalization is necessary prior to playback in line with findings of (Favrot and Buchholz, 2010) and (Zotter and Frank, 2019). Commonly, such equalization for Ambisonics recordings involves manually boosting mid and high frequencies based on the musical content (Pfanzagl-Cardone, 2023). Therefore, an additional advantage of the proposed approach is that it not only compensates for the acoustics of the reproduction room but also implicitly corrects the spectral characteristics of the Ambisonics recordings themselves.

The loudspeaker experiment demonstrated the robustness of the proposed algorithm against listener head rotation and displacement from the sweet spot, located at the center of the loudspeaker array. Notably, the approach showed improved performance for real recordings of piano, violin, and voice in the theater environment. This improvement was achieved despite mismatches between the optimization scenario, in which loudspeakers were used to measure RIRs, and the real recording scenario. These mismatches include differences in source positions and source widths, particularly for the piano, which has a significantly broader spatial extent than a loudspeaker. This highlights the effectiveness of the energy-based compensation strategy employed in our method.

The results from both headphone and loudspeaker experiments demonstrated that using only two Ambisonics channels (the WY configuration) with the proposed compensation approach is sufficient to outperform conventional reproduction of Eigenmike recordings in both anechoic and reverberant environments. From a data transmission perspective, this configuration requires only three signals: the beamformed direct signal, along with the W and Y spherical harmonic components. Compared to conventional 4th order Ambisonics, which requires 32 channels, this represents a substantial reduction in both storage and transmission requirements. Comparison between the compensated cases showed that adding one more Ambisonics channel, as in the WYX configuration, can improve both overall quality and spatial naturalness, potentially increasing the robustness of the compensation approach. The use of complete 1st order Ambisonics (WYXZ) did not show a significant improvement in mean scores compared to WYX, suggesting no clear perceptual advantage. Additionally, employing all available Ambisonics channels in the WYXZR configuration did not result in noticeable improvements over WYX or WYXY. Overall, the findings suggest that the truncated 1st order Ambisonics configuration (WYX) offers a favorable balance between the number of channels used and the quality of the reproduced sound.

-A primary limitation of the proposed method is that the $T_{60}$ of the reproduction room must be lower than or at most similar to that of the recording room. However, this constraint is generally not problematic, as it reflects typical real-world scenarios. For example, a concert recorded in a reverberant space is often reproduced in a typical room or studio listening environment. A second limitation relates to the optimization procedure, which requires impulse responses from the recording room. Experimental results with real instruments suggest that a rough estimation of the RIRs in the recording environment may be sufficient for effective optimization, though this requires further investigation. Additionally, this study focused exclusively on single-source recording and reproduction. Future work will involve extending the method to support reproduction of multiple simultaneous sound sources.

### ACKNOWLEDGMENTS

This work was supported by DFG Sonderforschungsbereiche (ID: 352015383–SFB 1330 C2, HAPPAA). The AI-based tool ChatGPT (OpenAI) was used to assist with language editing during the preparation of this manuscript.

## Author Declarations

## Conflict of Interest

The authors declare no conflicts of interest related to the content of this article.

## Ethics Approval

The listening tests conducted in this study were approved by the appropriate ethics committee at Carl von Ossietzky University of Oldenburg, and all participants provided informed consent in accordance with institutional guidelines.

## Data Availability

The data that support the findings of this study are available from the corresponding author upon request. Audio examples, simulation scripts, and evaluation materials used in the listening tests can be shared for academic research purposes.